\documentclass[12pt]{article}
\usepackage{hyperref}
\usepackage{sint_modi,cite}
\usepackage{amsmath,amssymb}
\usepackage{epsfig}
\usepackage{amssymb}
\usepackage{graphics} 
\usepackage{graphicx}
\usepackage{epsfig}

\usepackage{amssymb}
\usepackage{epsfig}

\def\a{\alpha}
\def\b{\beta}
\def\g{\gamma}
\def\G{\Gamma}
\def\D{\Delta}
\def\d{\delta}

\def\s{\sigma}
\def\S{\Sigma}

\def\beq{\begin{equation}}
\def\eeq{\end{equation}}
\def\beqn{\begin{eqnarray}}
\def\eeqn{\end{eqnarray}}
\def\ba{\begin{eqnarray}}
\def\ea{\end{eqnarray}}

\def\m{{\tt -}}

\def\xprim2bar{\overline{x}^{\prime\prime}}

\def\beq{\begin{equation}}
\def\eeq{\end{equation}}
\def\tr{{\bf tr}}

\newcommand{\beqa}{\begin{eqnarray}}
\newcommand{\eeqa}{\end{eqnarray}}

\let\a=\alpha   \let\b=\beta   \let\g=\gamma   \let\d=\delta
         
        \let\m=\mu
\let\n=\nu                 
\let\s=\sigma        

\let\G=\Gamma  \let\D=\Delta   
         \let\S=\Sigma

\newcommand{\Real}{{\rm Re}\,}
\newcommand{\re}{{\rm e}}

\newcommand{\rD}{{\rm D}}
\newcommand{\Seff}{S_{\rm eff}}

\newcommand{\oV}{\overline V}
\newcommand{\oH}{\overline H}
\newcommand{\oh}{\overline h}
\newcommand{\ov}{\overline v}
\newcommand{\dpp}{\d^{(4)}_{p'p''}}

\let\a=\alpha   \let\b=\beta   \let\g=\gamma   \let\d=\delta
         
        \let\m=\mu
\let\n=\nu                 
\let\s=\sigma        

\let\G=\Gamma  \let\D=\Delta   
         \let\S=\Sigma  

\newcommand{\be}{\begin{equation}}
\newcommand{\ee}{\end{equation}}
\newcommand{\bea}{\begin{eqnarray}}
\newcommand{\eea}{\end{eqnarray}}

\usepackage{graphics}
\usepackage{graphicx}
\def\tr{{\rm tr}}
\newcommand{\eq}[1]{Eq.~(\ref{#1})}
\newcommand{\fig}[1]{Fig.~\ref{#1}}

\newcommand{\sect}[1]{Section~\ref{#1}}

\def\A5{(A_5)_{\rm lat}}
\def\thintablerule{\hrule height0.4pt}
\begin{document}

\begin{flushright}
\small{
WUB/12-14
}
\end{flushright}

\vskip 1.5cm
\centerline{\LARGE Mean-Field Gauge Interactions in Five Dimensions II.}
\vskip 0.3cm
\centerline{\LARGE  The Orbifold.}

\vskip 2 cm
\centerline{\large Nikos Irges$^1$, Francesco Knechtli$^2$ and Kyoko Yoneyama$^2$}
\vskip1ex
\vskip.5cm
\centerline{\it 1. Department of Physics}
\centerline{\it National Technical University of Athens}
\centerline{\it Zografou Campus, GR-15780 Athens Greece}
\vskip .4cm
\centerline{\it 2. Department of Physics, Bergische Universit{\"a}t Wuppertal}
\centerline{\it Gaussstr. 20, D-42119 Wuppertal, Germany}
\begin{center}
{\it e-mail: irges@mail.ntua.gr, knechtli@physik.uni-wuppertal.de,\\
yoneyama@physik.uni-wuppertal.de}
\end{center}
\vskip 1.5 true cm
\thintablerule
\vskip 2.0ex
\leftline{\bf Abstract}
\vskip 1.0ex\noindent
We study Gauge-Higgs Unification in five dimensions on the lattice by means of the
mean-field expansion. We formulate it for the case of an $SU(2)$ pure gauge theory and
orbifold boundary conditions along the extra dimension, which explicitly break the
gauge symmetry to $U(1)$ on the boundaries. Our main result is that the gauge boson mass
computed from the static potential along four-dimensional hyperplanes is nonzero implying
spontaneous symmetry breaking. This observation supports earlier data from 
Monte Carlo simulations \cite{NFMC}.
\vskip 2.0ex
\thintablerule

\vskip-0.2cm
\newpage

\section{Introduction}

The phase diagram of five-dimensional gauge theories is surprisingly rich.
On an infinite, hypercubic, anisotropic lattice it is parametrized by the two dimensionless parameters
$\b_4=\b/\g$ and $\b_5=\b \g$, where $\b$ is the lattice coupling and 
$\g$ the anisotropy parameter. In part I of this work \cite{MFtorus} we explored the phase diagram
of a five-dimensional $SU(2)$ gauge theory with fully periodic boundary conditions,
using an expansion around a mean-field background \cite{DZ}. We concentrated on the regime
where $\b\sim O(1)$ and $\g<1$ where a line of second order phase transitions was 
observed. In the vicinity of this phase transition the system reduces dimensionally to four
dimensions and in the continuum limit the physics is consistent with what one would expect 
from the lightest states with four dimensional quantum numbers. For the fully periodic system
this would be a four dimensional $SU(2)$ gauge theory coupled to an adjoint scalar 
in the confined phase \cite{NFconfinement}. 

In this work we extend the construction of \cite{MFtorus}
by changing the boundary conditions in the fifth dimension from periodic to orbifold. 
The embedding of the orbifold projection in the geometry introduces boundaries
at the "ends" of the fifth dimension (which is now an interval) and its embedding into the gauge group
as well known by now \cite{HMR}, alters the field content surviving on the boundaries.
In this respect, for $SU(N)$ there are at least two possibilities.  
One is when the orbifold action is such that the adjoint set of scalars 
(i.e. the extra-dimensional components of the gauge field)
is projected out at the boundaries and one is left there with just a pure $SU(N)$ gauge theory.
This construction allows one to carry out an analysis similar to \cite{NFconfinement} but in a context that is 
directly generalizable to QCD once fermions are also added. 
The other possibility is to use the orbifold action to
project out some of the gauge fields and some of the scalars. With such a choice it is possible to realize
a field content similar to the one of the bosonic sector of the Standard Model. At a first stage we 
consider an $SU(2)$ bulk gauge group. The orbifold then leaves a $U(1)$ theory
coupled to a complex scalar on the boundaries. This setup could serve as the simplest prototype 
of the Higgs mechanism.

The idea that the Standard Model Higgs particle may be the remnant of an extra
dimensional gauge field is not new \cite{Hosotani}.
Also, investigations of five-dimensional gauge theories with a lattice regularization
have both an analytical and a Monte Carlo past.
The analytical work has been concentrated around the question of the existence or not of a layered
phase \cite{5Dan} and around the existence or not of an ultraviolet fixed point \cite{5DFP}.
Monte Carlo simulations have been mostly looking for the layered phase \cite{LPMC},
for first order bulk \cite{1PTMC} (the pioneering work in this direction) 
or second order \cite{2PTMC} (finite temperature and bulk) phase transitions and
dimensional reduction via localization \cite{FrancAM}. 
All of these lattice investigations have been carried out on fully periodic lattices.
Recently there has been interest also in lattices with orbifold boundary conditions
\cite{NFMC}, \cite{MM}. 
Our motivation to look more carefully at the phase diagram of five-dimensional
orbifold gauge theories stems from the fact that
Coleman-Weinberg computations \cite{ABQ}, continuum perturbation theory at
one loop \cite{OrbPert} and exploratory lattice Monte Carlo simulations \cite{NFMC}
indicate that once a Higgs mass is generated by quantum effects, 
it seems to remain finite, despite the non-renormalizable nature of the higher dimensional theory.

A first attempt to probe analytically the regime away from the perturbative point, in order to see if there is 
a dynamical mechanism of spontaneous symmetry breaking (SSB) triggered by the gauge field
(without the presence of fermions) was made in \cite{NFM}.  The idea there was to use the 
$SU(N)$ Symanzik lattice effective action \cite{Symanzik} ($a$ is the lattice spacing and
$N_5$ the number of lattice points in the fifth dimension)
\bea
-{\cal L}_{\rm Sym} & = & \frac{\b}{4Na}{\rm tr}\{F\cdot F\}+
\sum_{p_i}{c^{(p_i)}(N_5,\b)}\;{a^{p_i-4}}\;{\cal O}^{(p_i)}+\ldots
\label{efflagran}
\eea
which is determined by a finite number of dimensionless coefficient 
functions $c^{(p_i)}(N_5,\b)$ on an infinite spatial isotropic lattice, provided that one can
consistently truncate the expansion. The ansatz in \cite{NFM} was to truncate the expansion after the
first two higher dimensional operators: one of order $a$ corresponding 
to the lowest order dimension five boundary counterterm
parametrized by the coefficient function $c^{(5)}(N_5,\b)$ and 
one of order $a^2$ corresponding to the dimension six
bulk operator ${\rm tr} (DF)^2$ parametrized by the coefficient 
function $c^{(6)}(N_5,\b)$. 
To extract information about SSB in this setup, one assumes a vacuum expectation 
value (vev) $v$ for one of the  $A_5$ components of the five-dimensional gauge field $A_M$
and uses the truncated expansion expanded around this vev to compute a Coleman-Weinberg type potential 
$V(\a)$ for the dimensionless quantity
\be
\a = \frac{g_5R}{\sqrt{2\pi R}}v = 
     \sqrt{\frac{NN_5}{\beta\gamma}}\frac{a_4v}{\pi} \, ,
\ee
with $g_5$ the five-dimensional coupling, $\pi R=N_5a_5$ the size of the fifth dimension and 
we have written this formula for an anisotropic lattice with
lattice spacings $a_4$ and $a_5$ ($\gamma=a_4/a_5$ in the classical limit).
One then finds the preferred value for
$\a$ by minimizing $V(\a)$ \cite{ABQ,Kubo:2001zc} and calls it $\a_{\rm min}$.
As a consequence, the otherwise massless gauge boson develops a mass due to this vev equal to 
\be
m_Z = \frac{\a_{\rm min}}{R} \,. \label{zmass}
\ee
This is the Hosotani mechanism, applied to the case of the orbifold.
The result of the analysis of \cite{NFM}, performed at $\g=1$, was that indeed there exist values of $c^{(5)}$ and $c^{(6)}$ that yield 
a "Mexican hat" Higgs potential that triggers SSB with the Higgs particle having a mass of similar order as the gauge boson.
In particular, it was shown that a non-zero $c^{(6)}$ is able to trigger SSB by itself, by shifting $\a_{\rm min}$ from 
integer (for which there is no SSB) to half integer. This could be the main 
phenomenological gain from the complications encountered by entering in the interior of the phase diagram,
in view of the fact that in the conventional continuum approaches 
where one takes $c^{(p_i)}(N_5,\b)=0$, one necessarily needs fermions in order to trigger
SSB \cite{Kubo:2001zc} (i.e. a non-integer $\a_{\rm min}$) and even
if SSB is achieved, the Higgs typically turns out to be generically too light \cite{SSS}. 
In the absence of a non-perturbative control of the theory, in \cite{NFM} the coefficients $c^{(5,6)}$
were treated as free parameters. Non-perturbatively however they are not free parameters and 
it is not guaranteed that the quantum theory generates
values for the coefficients that trigger SSB.

In this work we improve on these approximations by computing the Wilson loop 
and the mass spectrum of the lightest states, in the mean-field expansion
far from the five-dimensional perturbative point, close to the bulk phase transition.
The formalism involved is very similar to the one developed in \cite{MFtorus}
and therefore will be heavily used. 
We show that the mean-field expansion predicts the spontaneous breaking of the boundary gauge symmetry 
already in the pure gauge system and allows for a Higgs like scalar of similar mass as the mass of the broken gauge field.
With infinite four-dimensional lattices, the parameters of our model are $\b$, $\g$ and $N_5$.
To the extent that the mean-field expansion is a good description of the non-perturbative system,
any result stemming from this approach should be taken seriously.
In fact, the first exploratory Monte Carlo studies of the orbifold theory \cite{NFMC} had earlier reached similar conclusions. 

In \sect{s_mfform} we give a short review of the mean-field expansion formalism.
In \sect{s_orbmf} we apply the general formalism to
the five-dimensional $SU(2)$ lattice gauge theory with orbifold boundary conditions along the extra dimension.
In \sect{s_ssb} we present our numerical results and in \sect{s_concl} our conclusions. In the Appendices we detail the mean-field
calculations of the propagator with orbifold boundary conditions and of the mass spectrum.
 
\section{A short review of the mean-field formalism \label{s_mfform}}

The partition function of a gauge theory on the lattice is
\be
Z = \int \rD U \re^{-S_W[U]} \,,\qquad
S_W[U]=\frac{\beta}{2N}\sum_{p}\Real\tr\{1-U(p)\} \,,
\ee
where $S_W[U]$ is the Wilson plaquette action and $p$ denotes oriented
plaquettes (i.e. each plaquette is counted with two orientations). 
In the mean-field approach \cite{DZ}
the link variables $U$ are traded for the complex quantities $H$ and $V$ and in terms of these 
one rewrites the partition function as
\be
Z = \int \rD V \int \rD H \, \re^{-\Seff[V,H]} \,,\quad
\Seff = S_W[V] + u(H) + (1/N)\Real {\rm tr} \{HV\} \,, \label{effaction}
\ee
where the effective mean-field action $u(H)$ is defined via
\be
\re^{-u(H)} = \int \rD U \, \re^{(1/N)\Real\tr\{UH\}} \,. \label{ueff}
\ee
The mean-field or zeroth order approximation amounts to finding the minimum
of the effective action when
\bea
H\longrightarrow \bar{H}\mathbf{1} \,,&
V\longrightarrow \bar{V}\mathbf{1} \,,&
\Seff[\bar{V},\bar{H}]\;\mbox{=minimal} \,. \label{saddlepoint}
\eea
The zeroth order saddle point solution or "mean-field background"
$[\oV]$ can be easily obtained by taking 
derivatives of \eq{effaction} with respect to $V$ and $H$
and require them to vanish. One then has
\be
{\overline V} = -\frac{\partial u}{\partial H}\Biggr|_{{\overline H}}\, ,\hskip 1cm
{\overline H} = -\frac{\partial S_W[V]}{\partial V}\Biggr|_{\overline V}\, .
\label{mf0}
\ee
The above two equations are the ones that make the action extremal and define the 
mean-field solution to zeroth order.  
The free energy per lattice site is
\be
F = -\frac{1}{\cal N}\ln (Z).
\ee
At 0'th order we simply have
\be
F^{(0)} = \frac{S_{\rm eff}[{\overline V},{\overline H}]}{\cal N}.
\ee
Gaussian fluctuations are defined by setting
\bea
H = \bar{H} + h \;& \mbox{and}\; & V = \bar{V} + v \, .
\eea
We impose a covariant gauge fixing on $v$. In \cite{Ruhl:1982er} it was shown
that this is equivalent to gauge-fix the original links $U$.
The integral
\be
z = \int \rD v \int \rD h \, \re^{-S^{(2)}[v,h]} = \frac{(2\pi)^{|h|/2}(2\pi)^{|v|/2}}
{\sqrt{{\rm det}[(-{\bf 1}+{K}^{(hh)}{K}^{(vv)})]}}
\ee
introduces the pieces of the propagator
\bea
&& \left.\frac{\delta^2\Seff}{\delta H^2}\right|_{\oV,\oH}h^2 = 
h_iK^{(hh)}_{ij}h_j = h^T K^{(hh)} h\\
&&  \left.\frac{\delta^2\Seff}{{\delta V}{\delta H}}\right|_{\oV,\oH}v h = 
v_iK^{(vh)}_{ij}h_j = v^T K^{(vh)} h\\
 &&\left.\frac{\delta^2\Seff}{{\delta V^2}}\right|_{\oV,\oH}v^2 = 
v_iK^{(vv)}_{ij}v_j = v^T K^{(vv)} v
\eea
which will be used extensively later. The quadratic part of the effective action
is $S^{(2)}[v,h]=\frac{1}{2}\left(
h^T K^{(hh)} h + 2v^T K^{(vh)} h + v^T K^{(vv)} v\right)$, $|h|$ and $|v|$ denote
the dimensionalities of the fluctuation variables $h$ and $v$.

We would like to compute the expectation value of observables
\be
\langle {\cal O} \rangle  = 
\frac{1}{Z} \int \rD U \, {\cal O}[U] \re^{-S_W[U]} 
\ee
in the mean-field expansion. To first order it is given by the formal expression \cite{MFtorus}
\be
\langle {\cal O} \rangle  =
{\cal O}[{\overline V}] + \frac{1}{2}{\rm tr} 
\left\{\frac{\delta^2{\cal O}}{\delta V^2}\Biggr|_{\oV} K^{-1}\right\} \, ,\label{correction}
\ee
with
\be
K=-K^{(vh)}{K^{(hh)}}^{-1}K^{(vh)}+K^{(vv)}+K^{({\rm gf})}\label{Kdef}
\ee
and the second derivative of the observable is taken in the mean-field background.
$K^{({\rm gf})}$ is the contribution from the gauge fixing term.
$K^{(vh)}$ actually turns out to be proportional to the unit matrix and drops out from all expressions.
The free energy at this order becomes
\be
F^{(1)} = F^{(0)} - \frac{1}{\cal N}\ln(z).
\ee
To extract the mass spectrum, we denote a generic, gauge invariant, time dependent 
observable as ${\cal O} (t)$ and its connected version as ${\cal O}^c (t)= {\cal O} (t_0+t){\cal O} (t_0) $.
Defining the correlator
\be
C (t) = <{\cal O}^c (t) > -
<{\cal O} (t_0+t)> <{\cal O} (t_0)>\, ,
\ee
to first order in the fluctuations the expression reduces to $C(t)=C^{(1)}(t)$ with
\be
 C^{(1)} (t) = \frac{1}{2} {\rm tr} 
\left\{ \frac{\d^{(1,1)} {\cal O}^c (t)}{\d^2 V } K^{-1}\right\} \, ,
\label{Higgsmass1}
\ee
where the notation $\d^{(1,1)}$ means one derivative acting on each 
of the ${\cal O}(t_0+t)$ and ${\cal O}(t_0)$. The mass of the lowest lying state is then
\be
m = \lim_{t\to \infty} \ln \frac{C^{(1)} (t)}{C^{(1)} (t-1)}\, .
\ee
It turns out that in order to extract the mass of the vector one needs to go to second
order in the mean-field expansion.
Physical expectation values are formally given at this order by \cite{MFtorus}
\bea
\langle {\cal O}\rangle  &=& {\cal O}[\overline V]+  \frac{1}{2}\left(\frac{\d^2  {\cal O}}{\d
  V^2}\right)_{ij}\left(K^{-1} \right)_{ij} \nonumber\\
&+&\frac{1}{24}\sum_{i,j,l,m}\left(\frac{\d^4  {\cal O}}{\d
  V^4}\right)_{ijlm}\Bigl( (K^{-1})_{ij}
(K^{-1})_{lm}+(K^{-1})_{il}(K^{-1})_{jm}+(K^{-1})_{im}(K^{-1})_{jl}\Bigr). \nonumber\\
\label{msecond}
\eea
To extract the mass from the connected correlator is straightforward. Again, all time
independent contribution (self energies) cancel from connected correlators and
at the end the mass is obtained from 
\be
m = \lim_{t\to \infty} \ln \frac{C^{(1)} (t)+C^{(2)} (t)}{C^{(1)} (t-1)+C^{(2)} (t-1)}\label{mcor2}
\ee
where $C^{(2)} (t) $ is the next to leading order correction to the $Z$-boson correlator.

\section{The lattice orbifold in the mean-field expansion \label{s_orbmf}} 

We will now apply the formalism we described in general terms to a
specific example: an $SU(2)$ lattice gauge theory in 5 dimensions with Dirichlet boundary
conditions for certain components of the gauge field along the fifth dimension. 

The discretized version of the $S^1/\mathbb{Z}_2$ orbifold defined on a five-dimensional
Euclidean lattice was constructed in \cite{NForbdef}. 
The points on the lattice are labeled by integer coordinates $n\equiv\{n_M\}$ but
we will often use the notation $n_0=t$ for the time component.
The periodic spatial directions ($M=\m=1,2,3$) have dimensionless extent $L=l/a_4$ and
the time-like direction ($M=0$) has extent $T$. 
The fifth dimension ($M=5$) has extent $N_5=\pi R/a_5$. 
The gauge-unfixed anisotropic mean-field Wilson plaquette action reads 
\bea
S_{\rm eff} &=&
-\frac{\b_4}{2}\sum_{n_\m}\sum_{n_5=1}^{N_5-1}\Biggl[\sum_{\m<\n}{\rm Re }\; 
{\rm tr}\; V_{p\notin {\rm bound}}(n; \m,\n)\Biggr] \nonumber \\
&& -\frac{\b_5}{2}\sum_{n_\m}\sum_{n_5=0}^{N_5-1}\Biggl[\sum_{\m}{\rm Re }\; 
{\rm tr}\; V_{p\notin {\rm bound}}(n; \m,5)\Biggr] \nonumber \\
&& -\frac{\b_4}{4} \sum_{n_\m}\Biggl[ \sum_{\m<\n}\sum_{n_5=0,N_5}{\rm Re }\; 
{\rm tr}\; V_{p\in {\rm bound}}(n; \m,\n)\Biggr] \nonumber\\
&& +\sum_{n_\m}\sum_{n_5=1}^{N_5-1}\sum_\m\left[ u_2(\rho(n,\m)) + 
\sum_\a h_\a(n,\m) v_\a(n,\m) \right]\nonumber\\
&& +\sum_{n_\m}\sum_{n_5=0}^{N_5-1}\left[ u_2(\rho(n,5)) + 
\sum_\a h_\a(n,5) v_\a(n,5) \right]\nonumber\\
&& +\sum_{n_\m}\sum_\m\sum_{n_5=0,N_5}\left[ u_1(\rho(n,\m)) + 
\sum_\a h_\a(n,\m) v_\a(n,\m) \right] \,,
\eea
where the effective mean-field actions $u_1$ and $u_2$ are computed in
Appendix \ref{s_effa}.
The couplings are defined as
\be
\b_4 = \frac{2Na_5}{g_5^2}, \hskip .5cm \b_5=\frac{2Na_4^2}{g_5^2a_5} \label{beta} \, .
\ee
In this work we will parametrize the infinite anisotropic lattice by the parameters $\b$ and $\g$, where $\b_4=\b/\g$ and $\b_5=\b\g$. 
A gauge transformation acts on a bulk link as
\be
U(n,M)\longrightarrow \Omega^{(SU(2))}(n) U(n,M)\Omega^{(SU(2))\dagger}(n+{\hat M})
\ee
on a boundary link as
\be
U(n,M)\longrightarrow \Omega^{(U(1))}(n) U(n,M)\Omega^{(U(1))\dagger}(n+{\hat M})
\ee
and on a link whose one end is in the bulk and the other touches the boundary as
\be
U(n,M)\longrightarrow \Omega^{(U(1))}(n) U(n,M)\Omega^{(SU(2))\dagger}(n+{\hat M})\, .
\ee
One possibility is to derive the orbifold theory from its parent circle theory.
In this setup a general link satisfies the orbifold projection condition
\be
\G\; U(n,M) = U(n,M), \hskip 1cm \G = {\cal T}_g{\cal R}\label{projG}
\ee
where the reflection property about the origin of the fifth dimension is
\bea
{\cal R}\; U(n, \m) &=& U({\overline n}, \m)\nonumber\\
{\cal R}\; U(n, 5) &=& U^\dagger({\overline n}-{\hat 5}, 5)
\eea
with
\be
n=(n_\m, n_5), \hskip .5cm {\overline n}=(n_\m, -n_5).
\ee
The transformation property under group conjugation is
\be
{\cal T}_g U(n, M) = g\,U(n,M)\,g^{-1} \, .
\ee
The boundary conditions that the above projections imply at their fixed points
amount to Dirichlet boundary conditions for some of the links
at the orbifold boundary hyperplanes.
Consequently the gauge group variables at the boundaries are
restricted to the subgroup of $SU(N)$ invariant under group conjugation
by a constant $SU(N)$ matrix $g$ with the property that
$g^2$ is an element of the center of $SU(N)$. Only gauge transformations
that commute with $g$ are still a symmetry at the boundaries and 
thus there, the orbifold breaks explicitly the gauge group.
In general, a bulk group $G$ breaks by $g$ to an equal rank subgroup
$\cal H$ on the two boundaries \cite{HMR}.

For $SU(2)$, which is the gauge group of our focus, we will take $g=-i\s^3$. This means that
at the orbifold fixed points the gauge group $SU(2)$ is broken to
the $U(1)$ subgroup parametrized by $\exp(i\phi\s^3)$, where $\phi$
are compact phases. In the continuum limit this implies that
$A_\m^{3}$ (the ``$Z$ gauge boson'') and $A_5^{1,2}$ (the ``Higgs'') satisfy Neumann boundary
conditions and $A_\m^{1,2}$ and $A_5^{3}$ Dirichlet ones.

In the mean-field approach, we parametrize the fluctuating fields in the bulk as
\bea
V(m,M) &=& v_0(n,M) + i \sum_{A=1}^3 v_A(n,M)\s^A\, ,\nonumber\\
H(m,M) &=& h_0(n,M) - i \sum_{A=1}^3 h_A(n,M)\s^A\, .
\eea
The $\s^A$ are the Pauli matrices.
On the boundaries instead, we use the parameterizations
\bea
V(n,M) & = & v_0(n,M) + i v_3(n,M)\s^3 \,, \nonumber\\
H(n,M) & = & h_0(n,M) - i h_3(n,M)\s^3 \, .
\eea
with $v_{0,A}\in\mathbb{C}$.
For later convenience we define the line
\bea
l^{(n_5)}({t_0, {\vec m}}) &=& \prod_{m_5=0}^{n_5-1}V((t_0,{\vec m},m_5); 5) 
\eea
and introduce the matrices
\be
\s^\a = \{ {\bf 1},\; i\s^A\},\hskip 1cm  {\overline \s}^\a = 
\{ {\bf 1},\; -i\s^A\}, \hskip .5cm A=1,2,3\, .
\ee 
For the computation of the $Z$ and Higgs masses we first define the orbifold projected Polyakov loop 
\be
P^{(0)}{(t, {\vec m})}=l^{(N_5)}(t, {\vec m}) \, g \, l^{(N_5)\dagger}(t, {\vec m})\, g^\dagger \, ,
\label{orbiPolya}
\ee
satisfying $\G P^{(0)} = P^{(0)}$,
in terms of which we define the field $\Phi^{(0)}{(t, {\vec m})} = 
P^{(0)}{(t, {\vec m})} -P^{(0)\dagger}{(t, {\vec m})}$
and then the displaced Polyakov loop \cite{MFtorus}
\be
Z^{(0),A}_k(t, {\vec m}) = {\overline \s}^A \, V((t,{\vec
  m},0);k)\, \Phi^{(0)\dagger}{(t, {\vec m}+{\hat k})}\, V((t,{\vec m},0);k)^\dagger
\, \Phi^{(0)}{(t, {\vec m})}\, ,\label{gaugeboson}
\ee
which assigns a vector and a gauge index to the observable appropriate to a gauge boson.
The Higgs observable is derived from the averaged over space and time location
connected correlator
\be
{\cal O}^c_{H} (t) = \frac{1}{L^6 T} \sum_{t_0}\sum_{{\vec m}',{\vec
    m}''} {\rm tr} \{P^{(0)}(t_0, {\vec m}')\} {\rm tr} \{P^{(0)}(t_0+t, {\vec m}'')\}\label{con1}
\ee
and the $Z$-boson from the correlator
\be
{\cal O}^c_{Z} (t) = \frac{1}{L^6 T} \sum_{t_0}\sum_{{\vec m}',{\vec
    m}''} \sum_A \sum_{k} {\rm tr} \{Z^{(0),A}_k (t_0, {\vec m}')\}\,
{\rm tr} \{Z^{(0),A}_k (t_0+t, {\vec m}'')\}\, .\label{con2}
\ee
Out of the above defined objects one can straightforwardly extract the masses using the 
general results of the previous section.

\subsection{The mean-field background\label{s_background}}

In order to determine the background we need the effective potentials $u_1$ and $u_2$.
The effective potentials in the bulk are the same as on the torus, while on the boundaries they are
\be
u_1(H(n,M)) = -\ln(I_0(\rho))\,,\quad
\rho=\sqrt{(\Real h_0(n,M))^2+(\Real h_3(n,M))^2} \,.
\ee
For details see Appendix \ref{s_effa}.

Translation invariance along the dimensions $\mu=0,1,2,3$ means that we
can parametrize the saddle point solution, which minimizes $S_{\rm eff}$,
as follows \cite{Knechtli:2005dw}: for $n_5=0,1,\ldots,N_5$ (four-dimensional links)
\be
\oH(n,\mu)= \oh_0(n_5)\mathbf{1}\,,\qquad
\oV(n,\mu)= \ov_0(n_5)\mathbf{1}\,,\qquad
\forall n_\mu\,,\mu \,, \label{meanlinko4d}
\ee
and for $n_5=0,1,\ldots,N_5-1$ (extra-dimensional links)
\be
\oH(n,5)= \oh_0(n_5+1/2)\mathbf{1}\,,\qquad
\oV(n,5)= \ov_0(n_5+1/2)\mathbf{1}\,,\qquad
\forall n_\mu\,. \label{meanlinkoed}
\ee
The action at zeroth order reads (${\cal N}=L^3T{\cal N}_5$, ${\cal N}_5=N_5+1$)
\bea
&&\frac{\Seff[\oV,\oH]}{{\cal N}} = \frac{1}{{\cal N}_5}\Biggl\{
-\frac{\b_4}{2}(d-1)(d-2)
\left[  
\sum_{n_5=1}^{N_5-1}\ov_0(n_5)^4+\frac{1}{2}\ov_0(0)^4+\frac{1}{2}\ov_0(N_5)^4
\right]\nonumber \\
&&-\b_5(d-1)\sum_{n_5=0}^{N_5-1}\ov_0(n_5)(\ov_0(n_5+1/2))^2\ov_0(n_5+1)
\nonumber \\
&&+(d-1)\left[u_1(\oh_0(0))+u_1(\oh_0(N_5))+
\sum_{n_5=1}^{N_5-1}u_2(\oh_0(n_5))+\sum_{n_5=0}^{N_5}\oh_0(n_5)\ov_0(n_5)
\right] \nonumber \\
&&+\sum_{n_5=0}^{N_5-1}\left[u_2(\oh_0(n_5+1/2))+\oh_0(n_5+1/2)\ov_0(n_5+1/2)
\right]\Biggr\} \,.
\eea
The minimization equations lead to the following relations: for $n_5=0$
\bea
\ov_0(0)&=&-u_1'(\oh_0(0))=\frac{I_1(\oh_0(0))}{I_0(\oh_0(0))} \,,\\
\oh_0(0)&=&\b_4\left[(d-2)(\ov_0(0))^3+\g^2(\ov_0(1/2))^2\ov_0(1)\right] \,.
\eea
A prime on $u_1$ or $u_2$ denotes differentiation with respect to its argument.
Similarly, for $n_5=N_5$ we have
\bea
\ov_0(N_5)&=&-u_1'(\oh_0(N_5))=\frac{I_1(\oh_0(N_5))}{I_0(\oh_0(N_5))} \,,\\
\oh_0(N_5)&=&\b_4\left[(d-2)(\ov_0(N_5))^3+\g^2\ov_0(N_5-1)(\ov_0(N_5-1/2))^2
\right] \,.
\eea
For $n_5=1,\ldots,N_5-1$ (four-dimensional links)
\bea
\ov_0(n_5)&=&-u_2'(\oh_0(n_5))=\frac{I_2(\oh_0(n_5))}{I_1(\oh_0(n_5))} \,,\\
\oh_0(n_5)&=&\b_4\left[2(d-2)(\ov_0(n_5))^3
+\g^2\Bigl((\ov_0(n_5+1/2))^2\ov_0(n_5+1) \right. \nonumber \\
&&
\left.+\ov_0(n_5-1)(\ov_0(n_5-1/2))^2\Bigr)\right] \,.
\eea
For $n_5=0,\ldots,N_5-1$ (extra-dimensional links)
\bea
\ov_0(n_5+1/2)&=&-u_2'(\oh_0(n_5+1/2))=
  \frac{I_2(\oh_0(n_5+1/2))}{I_1(\oh_0(n_5+1/2))} \,,\\
\oh_0(n_5+1/2)&=&2\b_5(d-1)\ov_0(n_5)\ov_0(n_5+1/2)\ov_0(n_5+1) \,.
\eea

\subsection{Observables from fluctuations around the background}

In sect. 2 we described the general formalism for computing observables from fluctuations
around the mean-field background. Here we apply this formalism to our case and give the results
for the free energy, the scalar and vector masses. 

The lattice propagator, Fourier transformed along 
the spatial and time directions is an object that contains the information 
about the boundary conditions. We denote its components as
\be
{K}^{-1} = {K}^{-1}(p',n'_5,M',\a' ;p'',n''_5,M'',\a'')\, .
\ee
The momenta $p',p''$ are four dimensional momenta, however
we will further split the momenta into their time and spatial components:
$p=p_0,p_k$ whenever it is necessary, for clarity.
For a more detailed computation of the propagator we divert the reader at this point to Appendix B.  

\subsubsection{The free energy}

The free energy to first order is 
\bea
F^{(1)}&=&F^{(0)}+\frac{1}{2{\cal N}} \ln \Biggl[
\prod _{\a=0}^{3}{\rm det} \Bigl(\Upsilon (p',n_5',M',\a;p'',n_5'',M'',\a)
D_{\rm FP}^{({\rm orb})-2}\Biggr],
\label{FreeEnergy1}
\eea
where $D^{({\rm orb})}_{\rm FP}$ is the determinant of the Faddeev-Popov matrix, also computed in
Appendix B. The matrix $\Upsilon$ is defined as
\be
K^{-1}=\Upsilon^{-1}K^{(hh)}, \hskip 1cm 
\Upsilon=-{\bf 1} + K^{(hh)}({K}^{(vv)}+K^{({\rm gf})})\, .
\ee
There are torons both in the matrix $\Upsilon$ and $D_{\rm FP}^{({\rm orb})}$,
which can be regularized as on the torus \cite{MFtorus}, see the end of Appendix \ref{appb}.
 
\subsubsection{The Higgs and $Z$-boson masses}

For the Higgs, there is a non-trivial contribution already at first order. 
The result is
\be
C_H^{(1)}(t) = \frac{8}{{\cal N}^{(4)}}(P_0^{(0)})^2  \Pi^{(1)}_{\langle 1,1\rangle}(0,0) \,,
\label{Hcorr}
\ee
where $P_0^{(0)}$ is the Polyakov loop \eq{orbiPolya} evaluated on the background and
$\Pi^{(1)}_{\langle 1,1\rangle}(0,0)$ is defined in \eq{Pisymbols}.
The above correlator does not contain torons since the 00 component of the propagator does not contain any.

For the $Z$ the result is
\bea
&& C_{Z}^{(2)}(t)=\frac{4096}{({\cal N}^{(4)})^2}(P_0^{(0)})^4 (v_0(0))^4
\sum_{\vec p'}\sum_k\sin^2{p_k'}
  \Pi^{(2)}_{\langle 1,1\rangle}(1,1)^2\, , \label{Zcorr}
\eea
where $\Pi^{(2)}_{\langle 1,1\rangle}(1,1)$ is defined in \eq{Pisymbols}.
It contains regularizable torons, that is simultaneous zero modes in the propagator and the observable, 
whose contribution vanish in the infinite lattice volume limit.

Derivations can be found in Appendix \ref{s_appmass}.


\subsubsection{The static potential}

On the orbifold we have the three types of static potentials.
Here we will be interested in the potentials extracted from
Wilson loops in the four-dimensional hyperplanes,
along either one of the boundaries and in the middle of the orbifold
(i.e. at $n_5=N_5/2$).
We consider the Wilson loops of size $r$ along one of the 
three spatial dimensions and
we average over the possible orientations.
The exchange contribution (to $\d^2  {\cal O}^c/\d V^2$) at $n_5=0$ is
\bea
&&{\cal O}_{\rm ex}\equiv 
\frac{t^2}{L^3T}
2(\ov_0(0))^{2(t+n_3)-2} \d_{M'0}\d_{M''0} \nonumber \\
&&
\d_{n_50}(\d_{\a'0}\d_{\a''0}+\d_{\a'3}\d_{\a''3}) 
\d_{p_0'0}\d_{p_0''0}\left(\prod_{M=1,2,3}\d_{p_M'p_M''}\right)
\frac{1}{3}\sum_{k=1}^{3}2\cos{(p_k r)}
\; \d_{n_5'0}\d_{n_5''0} \nonumber\\\label{pot4d_39_orb}
\eea
and the self energy contributions
\bea
&&{\cal O}_{\rm se} \equiv 
\frac{t^2}{L^3T}
2(\ov_0(0))^{2(t+n_3)-2} \d_{M'0}\d_{M''0} \nonumber \\
&&
\d_{n_50}(\d_{\a'0}\d_{\a''0}-\d_{\a'3}\d_{\a''3})
\d_{p_0'0}\d_{p_0''0}\left(\prod_{M=1,2,3}\d_{p_M'p_M''}\right)
2\; \d_{n_5'0}\d_{n_5''0}\,.\label{pot4d_111_orb}
\eea
As for the torus \cite{MFtorus}, to first order we would like to compute
\bea
&&C_W^{(1)}=\frac{1}{2}\sum_{\a',\a''} \sum_{p_k'} \sum_{n_5',n_5''}\nonumber\\
&&{\cal O} \Bigl(0,p_k',n_5',0,\a';0,p_k',n_5'',0,\a''\Bigr)
{K}^{-1}
\Bigl(0,p_k',n_5',0,\a';0,p_k',n_5'',0,\a''\Bigr)\,, \label{bpotcorr}
\eea
where ${\cal O}={\cal O}_{\rm ex}+{\cal O}_{\rm se}$, which is to be substituted in the general expression for the first order corrected static potential
\be
V ={\rm const.} -\lim_{t\to \infty}\frac{1}{t}
\frac{C_W^{(1)}}{{\cal O}[{\overline V}]}\, .
\ee
Applied to the gauge boson exchange between two static charges on the boundary the general formula reduces to
\bea
&&V_4(0)=-\log({\ov_0(0)^2})-\frac{1}{2}\frac{1}{L^3T}
\frac{1}{(\ov_0(0))^2}\sum_{p_k'}\nonumber\\
&& \Biggl\{\frac{1}{3}\sum_k\Bigl[ 2\cos{(p_k'r)}+2\Bigr]
{K}^{-1} \left(0,p_k',0,0,0;0,p_k',0,0,0\right)\nonumber\\
&+& \frac{1}{3}\sum_k\Bigl[ 2\cos{(p_k'r)}-2\Bigr]
{K}^{-1}\left(0,p_k',0,0,3;0,p_k',0,0,3\right)\Biggr\}.
\label{potb}
\eea 

The formula for the static potential $V_4(N_5/2)$ for the static potential
along the four dimensional hyperplane in the middle of the orbifold is similar
to \eq{potb}, the difference being that the background and the propagator are 
evaluated at $n_5=N_5$ and in the last line of \eq{potb} there is a sum over 
all the gauge components $\alpha=1,2,3$.

\section{Spontaneous symmetry breaking \label{s_ssb}}

\subsection{The phase diagram \label{s_pd}}

\begin{figure}[!t]
\centerline{\epsfig{file=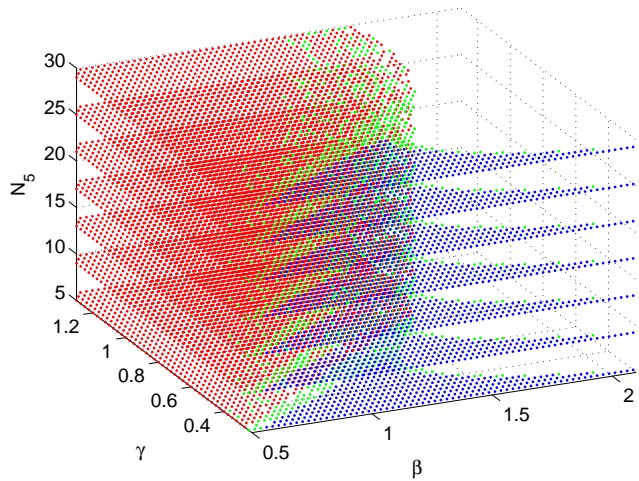,width=0.6\textwidth}}
\caption{\small The mean-field phase diagram of the $SU(2)$ orbifold theory
in the $(\b, \g, N_5)$ space. The color code is explained in the text.}
\label{f_phasediagram}
\end{figure}

Using the equations that determine the mean-field background in \sect{s_background},
one can extract the leading order approximation to the phase diagram.
The equations are solved iteratively and numerically.
The confined phase is defined as the phase where
$\ov_0(n_5)=0=\ov_{0}(n_5+1/2)$ for all $n_5$.
When $\ov_0(n_5)\ne 0$ and $\ov_0(n_5+1/2)=0$ for all $n_5$, we define the layered phase.
The Coulomb phase is defined where $\ov_0(n_5)\ne 0$ and $\ov_0(n_5+1/2)\ne 0$ for all $n_5$.
We do not find a phase where $\ov_0(n_5)= 0$ and $\ov_0(n_5+1/2)\ne 0$ for all $n_5$.
The background is sensitive to $\b, \g$ and $N_5$.
On \fig{f_phasediagram} we plot the phase diagram with color code,
red for the confined phase, blue for the layered phase and white for the Coulomb phase.
Green is used where for some reason the iterative process does not converge to a solution.
For the rest of this paper we will stay in the Coulomb phase.
%
%

As for the torus, the line that separates the Coulomb from the confined phase is of first order
for $\g$ larger than a value which is close to 0.7., below which it turns into a second order
phase transition. 
The order of the phase transition that the mean-field predicts must be taken
with care though, as a more careful, fully non-perturbative analysis should be done.

Next, using the quantities $C_W^{(1)}$, $C_Z^{(2)}$ and $C_H^{(1)}$ we analyze the physical
properties of the system on this phase diagram, with an emphasis on the issue of
spontaneous symmetry breaking (SSB).

\subsection{The Higgs}

\begin{figure}[!t]
\begin{minipage}{.5\textwidth}
\centerline{\epsfig{file=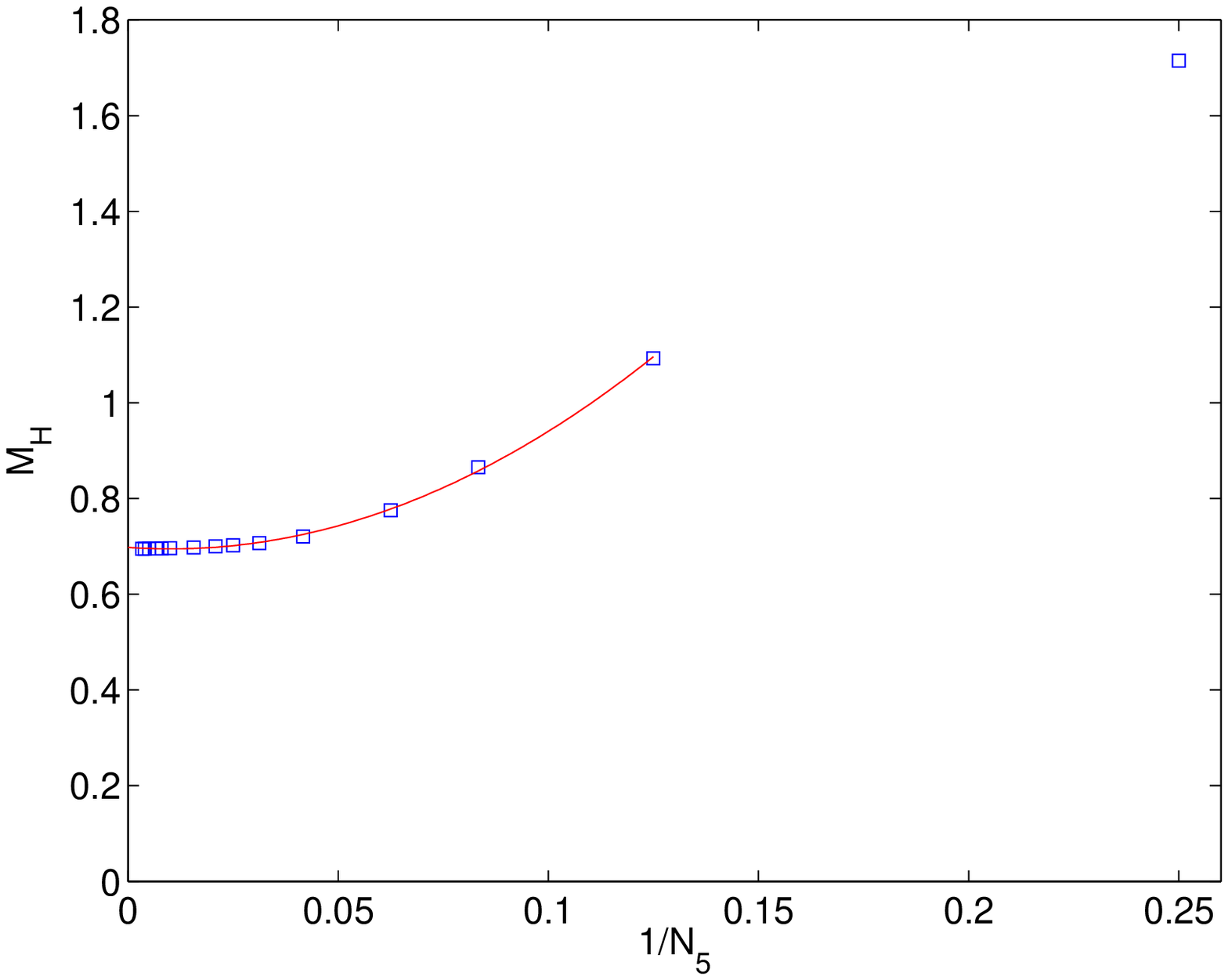,width=8cm}}
\end{minipage}
\begin{minipage}{.5\textwidth}
\centerline{\epsfig{file=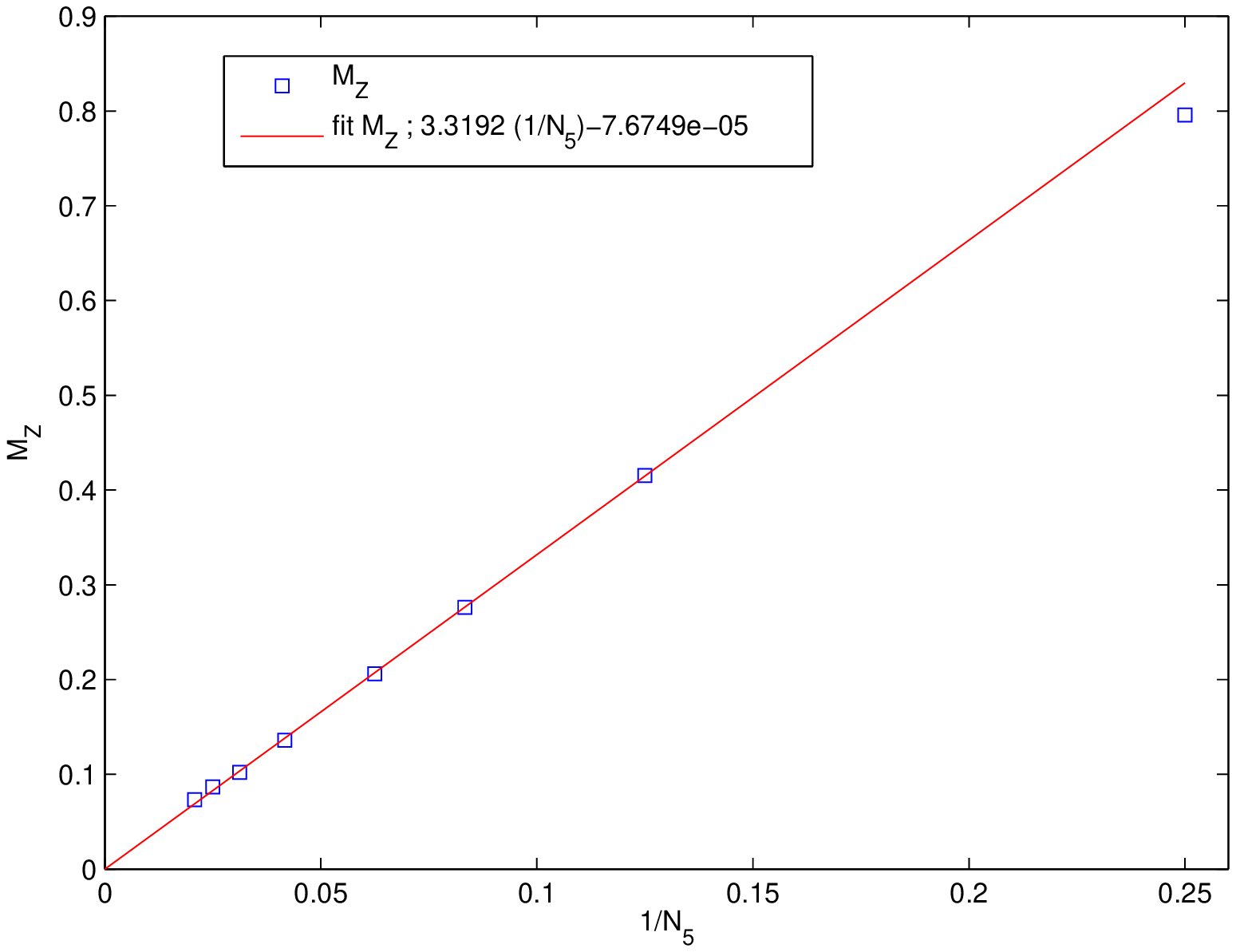,width=8cm}}
\end{minipage}
\caption{\small The Higgs mass $M_H$ (left plot) and the $Z$ boson mass $M_Z$
(right plot) as a function of $1/N_5$ at $\g=1$ for $\b=1.677$. The squares
are mean-field data, 
the line on the right plot is a quadratic fit and the line on the left plot
a linear fit.  
\label{f_masses_g1}}
\end{figure}

The Higgs mass in units of the lattice spacing $M_H=a_4m_H$, extracted from $C_H^{(1)}$
in \eq{Hcorr}, depends on $\b, \g$ and $N_5$.
The physical quantity of our interest is the Higgs mass in units of the radius of the fifth dimension
\be
F_1 = m_H\,R  = M_H\,\frac{N_5}{\g\,\pi} \, . \label{F1}
\ee
In perturbation theory, the one-loop result \cite{OrbPert} for $SU(N)$, expressed in lattice parameters 
(relevant for the isotropic lattice) is ($C_2(N)=(N^2-1)/(2N)$)
\be
M_H^{\rm pert.} = \frac{c\,\gamma\,\pi}{N_5^{3/2}\,\b^{1/2}}, \hskip .5cm 
c=\frac{3}{4\pi^2}\sqrt{N\,\zeta(3)\,C_2(N)} \, .\label{pertHiggs}
\ee
On the left plot in \fig{f_masses_g1} we show the $N_5$-dependence of $M_H$ for $\g=1$ at $\b=1.677$
near the phase transition.
We can see clearly that the perturbative formula is not valid at a generic point on the phase diagram.
The line on the left plot in \fig{f_masses_g1} is a quadratic fit. The phase transition is of first order,
which means that the mass in lattice units $M_H$ cannot be lowered to zero but approaches a
non-zero minimal value, which at $\b=1.677$ is approximately 0.69.
%
%

\subsection{The $Z$ boson}

The Wilson loop can decide if there is SSB.
We will choose our lattices so that the system is dimensionally reduced to four dimensions.
Then we can describe the boundary gauge theory in four-dimensional terms.
If the boundary $U(1)$ symmetry is spontaneously broken then the corresponding static potential 
extracted from $C_W^{(1)}$ \eq{bpotcorr}
should be fitted by a ($4d$) Yukawa form rather than by a Coulomb form. Starting from
\be
V_4(r) = -b \frac{e^{-m_Z r}}{r} + {\rm const.} \,,
\ee
where $b$ is a constant,
we define the quantity $y(r)= \log(r^2 F_4(r))$ where $F_4(r)={\rm d}\,V_4(r)/{\rm d}\,r$
from which we form the combination
\be
a_4 y'(r) = - M_Z + \frac{M_Z}{m_Z r +1}\, . \label{yprime}
\ee
By $M_Z=a_4m_Z$ we denote the $Z$ mass in lattice units.
We then determine $M_Z$ iteratively by requiring that a plateau for $-a_4y'(r) + M_Z/(m_Z r + 1)$
forms.
The plateaus, for large enough $L$, stabilize as $L$ is further increased, so that $M_Z$ 
at infinite $L$ depends on $\b$, $\g$ and $N_5$.
Note that in the case of a first order phase transition this corresponds to a $Z$ boson in
an infinite physical volume at a finite lattice spacing.

\subsubsection{Isotropic lattices}
 
On the right plot of \fig{f_masses_g1} we show the $M_Z$ plateau values as a function of $1/N_5$
at fixed $\b=1.677$ and $\g=1$, near the bulk phase transition.
The plateau values of $M_Z$ do not depend on $L$ for $L\ge200$ and there is no sign
of a plateau for a zero mass. 
A linear fit with slope $3.32$, which is very close to $\pi$, describes the data very well.
The most striking observation is that the boundary gauge boson is massive, pointing to the
dynamical spontaneous breaking of the $U(1)$ symmetry. Clearly, since $\b$ and $\g$ are kept fixed,
the masses on \fig{f_masses_g1} correspond
to different lattice spacings (the location of the phase transition $\beta_c$ depends on $N_5$).
Nevertheless, from the Kaluza--Klein description we expect to see an approximate $1/N_5$ dependence.
Comparing the data of \fig{f_masses_g1}
to \eq{zmass} our result indicates a value $\alpha_{\rm min}=1$. 
However the regime of Higgs masses in \fig{f_masses_g1}
corresponds to values of $F_1=m_H\,R\gg1$ which is not the regime
where the Coleman--Weinberg calculation is performed and for which $F_1<1$.
In perturbation theory
the minimum at $\a_{\rm min}=1$ is equivalent to
the one at $\a_{\rm min}=0$ and describes a situation without SSB. Clearly this is not the
case for the mean-field data.
%
\begin{figure}[!t]
\centerline{\epsfig{file=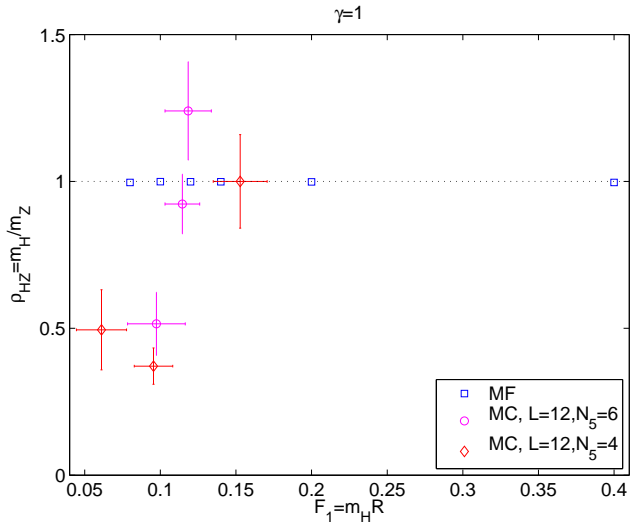,width=0.6\textwidth}}
\caption{\small The ratio of the Higgs to the $Z$ boson mass \eq{rho}.
Comparison of Monte Carlo (diamonds \cite{NFMC} and circles \cite{NFM})
and mean-field data (squares) at $\gamma=1$.}
\label{f_rho}
\end{figure}
%

On \fig{f_rho} we plot mean-field data (squares) for the ratio
\be
\rho_{HZ} = \frac{m_H}{m_Z} \,, \label{rho}
\ee
obtained using $N_5=4,6,8$ and $L=200$. As far as we have checked the
mean-field results in \fig{f_rho} are independent of $N_5$.
For $\g=1$ and $F_1$ in the range $[0.08,0.4]$ 
the Higgs and the $Z$ boson
are almost degenerate in mass and so $\rho_{HZ}\simeq1$.
As a result, $\a_{\rm min}\simeq F_1$ in this range of $F_1$ values.
Contrary to the data shown in \fig{f_rho}, on \fig{f_masses_g1},
where $\a_{\rm min}=1$, the values of
$F_1$ are larger than 2 and $\rho_{HZ}>2$.
We compare to the results from Monte Carlo simulations at $N_5=4$ (diamonds) \cite{NFMC} and
at $N_5=6$ (circles) \cite{NFM}, using $L=12$ and $T=96$.
There is good agreement between the mean-field data and the Monte Carlo data
on isotropic lattices, demonstrating that it is possible to obtain values
$\rho_{HZ}\ge1$.

We have also computed the $Z$ boson mass from the potential in the middle
of the bulk at $\gamma=1$. For $F_1=0.2$ ($N_5=8$), the mass $M_Z$ is
the same as the one extracted from the boundary potential. We find
$\rho_{HZ}\simeq1$ in the bulk for $F_1=0.2$ and $F_1=1.0$.

We find that the mass of the ground state extracted from the direct $Z$ correlator 
$C_Z^{(2)}$ in \eq{Zcorr} at this order in the mean-field expansion
does not depend on $\b$ or $\g$ and most importantly it
does not depend on $N_5$. This means that using this observable one can measure only its infinite 
$N_5$ limit value, which for finite $L$, turns out to be
\be
M_Z^{\rm dir.} = \frac{4\pi}{L}
\ee
like on the torus \cite{MFtorus}. This expression reflects the fact that this observable describes
two non-interacting gluons (that is why the $4\pi$). 
%
\begin{figure}[!t]
\begin{minipage}{.5\textwidth}
\centerline{\epsfig{file=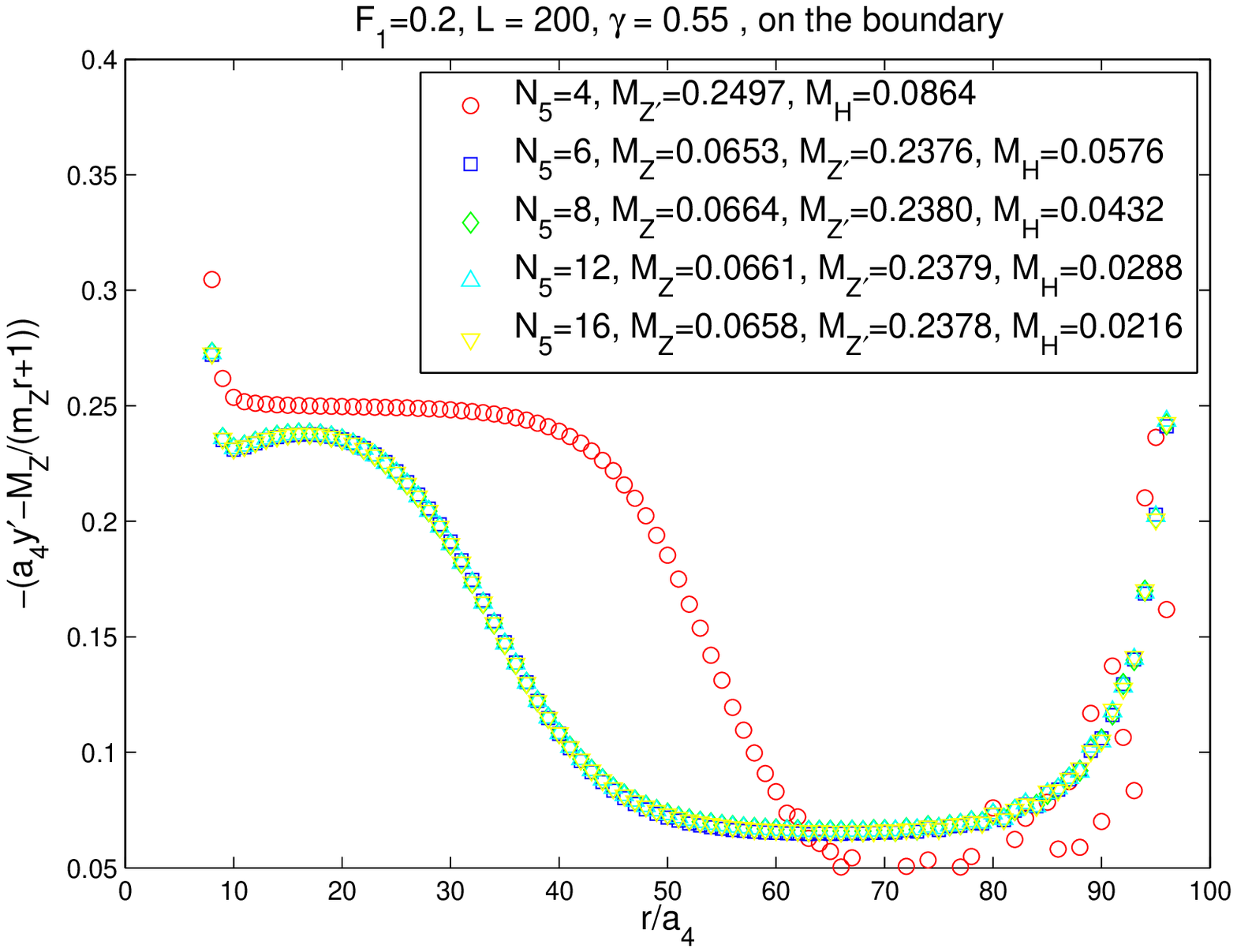,width=8cm}}
\end{minipage}
\begin{minipage}{.5\textwidth}
\centerline{\epsfig{file=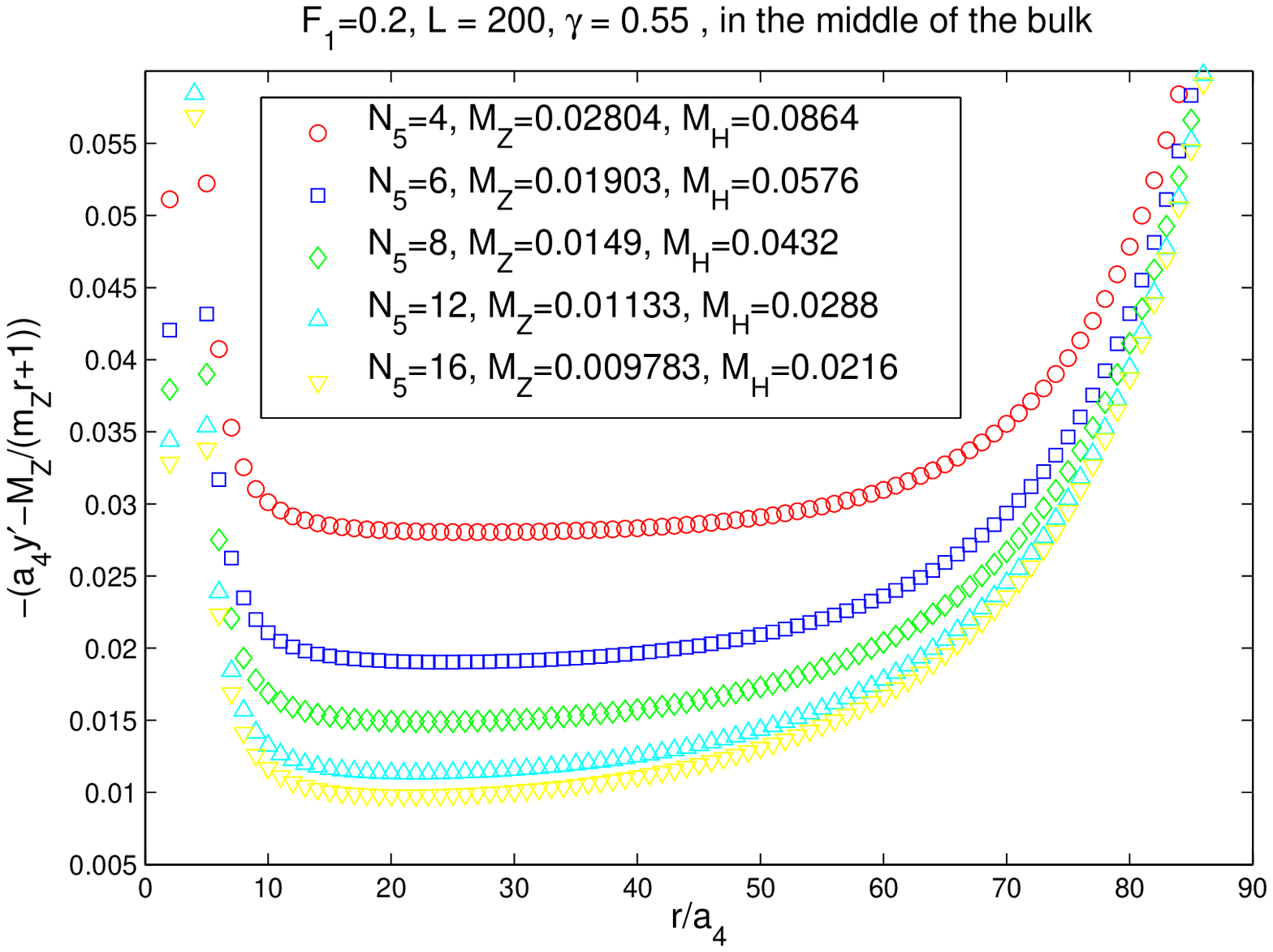,width=8cm}}
\end{minipage}
\caption{\small The combination $-a_4y'(r) + M_Z/(m_Z r + 1)$, cf. \eq{yprime} is plotted
for different values of $N_5$ at $\g=0.55$ and $F_1=0.2$ (for the boundary potential at $N_5=4$
we use $M_{Z'}$).
Boundary potential (left plot) and bulk potential (right plot).
\label{f_yukawa_g0p55}}
\end{figure}
%

\subsubsection{Anisotropic lattice}

Motivated by \sect{s_pd} we study SSB at $\gamma=0.55$ by extracting the Yukawa masses from
the static potential on the boundary and in the middle of the orbifold following \eq{yprime}.
As we vary $N_5$, we set $\beta$ to keep 
$F_1=0.2$ constant, which means that $M_H\propto1/N_5$, cf. \eq{F1}.

\begin{figure}[!t]
\begin{minipage}{.5\textwidth}
\centerline{\epsfig{file=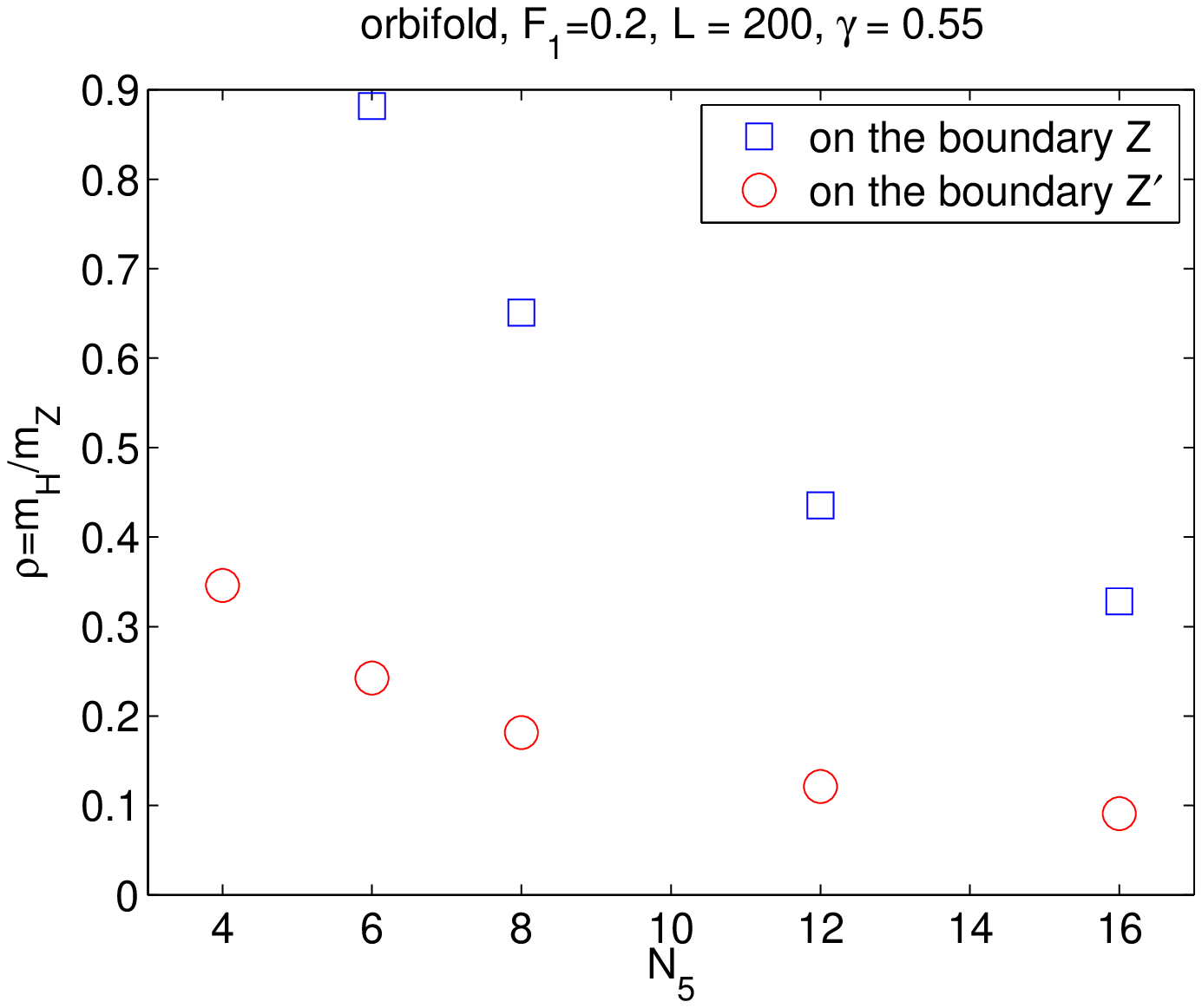,width=8cm}}
\end{minipage}
\begin{minipage}{.5\textwidth}
\centerline{\epsfig{file=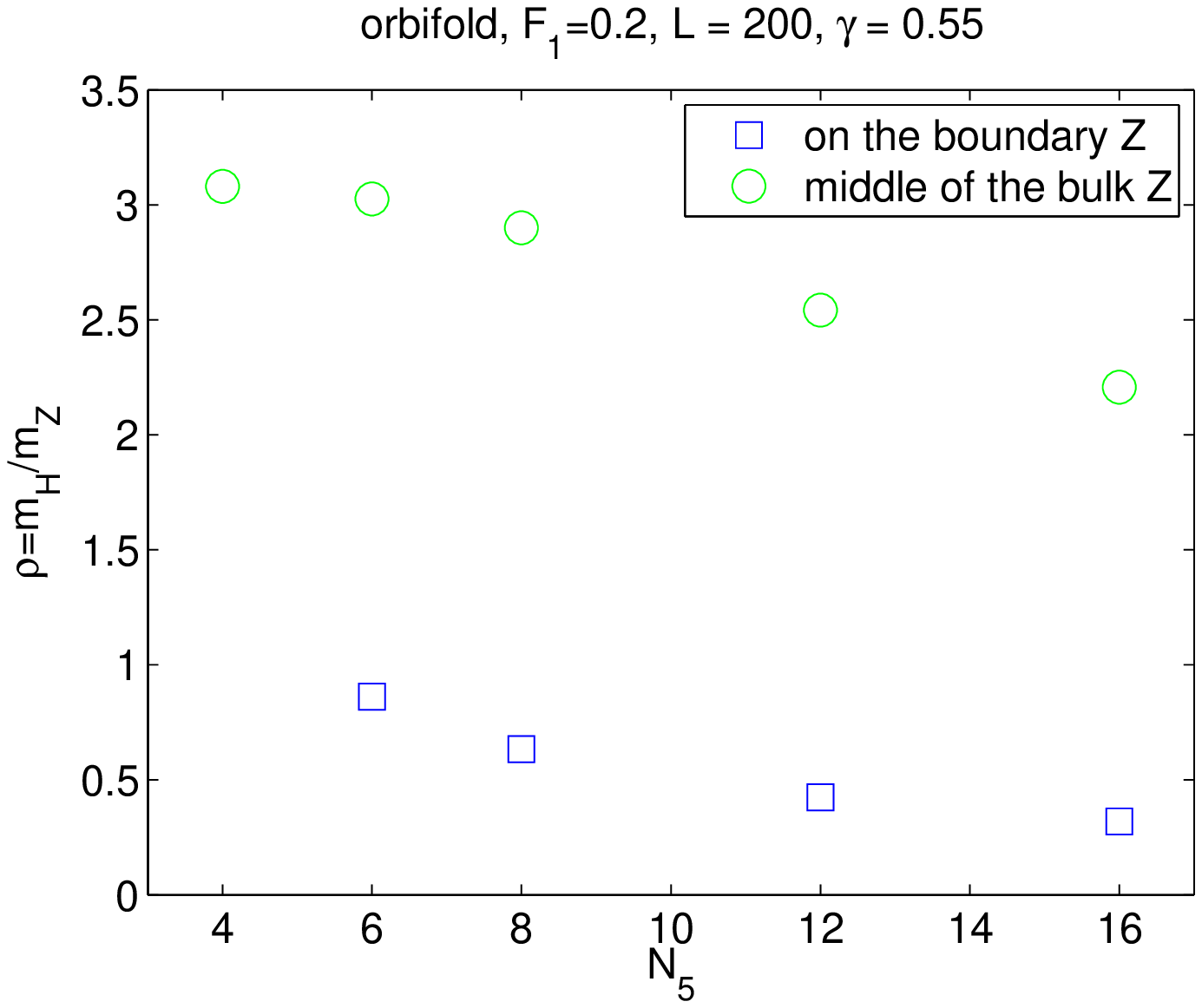,width=8cm}}
\end{minipage}
\caption{\small The ratio of the Higgs to the $Z$ boson mass \eq{rho} in
the mean-field extracted from the static potential. On the boundary (left
plot) and in the bulk (right plot). 
\label{f_rho_g0p55}}
\end{figure}

On the left plot in \fig{f_yukawa_g0p55} we show the plateaus for the boundary potential.
For $N_5\ge6$ there are two plateaus corresponding to masses $M_{Z'}>M_{Z}$ which do not
depend on $N_5$. 
At $N_5=4$ there is only one plateau whose value is very close to $M_{Z'}$ for $N_5\ge6$.
Therefore we identify the plateau at $N_5=4$ with the mass of a $Z'$ boson (although we do
not see the $Z$ boson state).
We checked that the Yukawa masses are independent of $L$ if $L$ is large.
For this we compared $L=200$ with $L=300$ and find no differences. These data establish
that the boundary theory is a spontaneously broken $U(1)$ theory and by comparing to
\eq{zmass} we get $\a_{\rm min}\approx0.039\,N_5$ for $N_5\ge6$. We find $\rho_{HZ}<1$
for $N_5\ge6$ and
this particular choice of parameters as shown on the left plot of \fig{f_rho_g0p55}.
Finally we remark that the boundary potential cannot be fitted by a four-dimensional
string-like fit.

On the right plot in \fig{f_yukawa_g0p55} we show the situation in the middle of the
orbifold. There is one plateau yielding a bulk $Z$ boson mass $M_Z$. $M_Z$ in the bulk
is decreasing as $N_5$ increases and is independent on $L$. This result indicates that
there is SSB also in the bulk, where we have $\rho_{HZ}>1$ as shown on the right
plot of \fig{f_rho_g0p55}. The situation on the orbifold is therefore completely different
than it is on the torus, where a Yukawa fit yields a mass $\approx 2\pi/L$ implying that
there is no SSB. This fact is by itself non trivial as it shows that the orbifold
boundary conditions are affecting the properties of the bulk. We observe a difference
between the Yukawa masses in the bulk as compared to those on the boundary. 
This situation is different than the one of the isotropic lattice, where we found the 
boundary and bulk Yukawa masses to be the same.
%
%
\section{Conclusions \label{s_concl}}

We have formulated the mean-field expansion in five dimensions on the lattice
for an $SU(2)$ gauge theory and orbifold boundary conditions. 
We have performed computations on the isotropic lattice and for anisotropy
parameter $\gamma=0.55$ and find that the
gauge boson mass extracted from the static potential along four-dimensional
hyperplanes is non-zero, both on the boundaries and in the middle of the orbifold.
The gauge boson mass does not depend on the spatial size of the lattice,
thus indicating dynamical spontaneous symmetry breaking. This result differs from the one
obtained in the perturbative limit of this theory, where the gauge boson remains
massless, but supports the first Monte Carlo simulations that were performed in \cite{NFMC}.


{\bf Acknowledgments.} We thank A. Kurkela, A. Maas and P. Weisz for discussions. 
K. Y. is supported by the Marie Curie Initial Training Network STRONGnet.
STRONGnet is funded by the European Union under Grant Agreement number 238353 
(ITN STRONGnet). N. I. thanks the Alexander von Humboldt Foundation for support.
N. I. was partially supported by the NTUA research program PEBE 2010.

\begin{appendix}
\section{Effective mean-field action \label{s_effa}}

The effective mean-field action is defined in \eq{ueff} as a function
of a gauge link $U\equiv U(n,M)$ and a complex matrix $H\equiv H(n,M)$.
On the orbifold we have to distinguish two cases, 
depending whether the link is a bulk $SU(2)$
link or a boundary $U(1)$ link.

\subsection{Bulk $SU(2)$ links}

In order to compute \eq{ueff}
for gauge group $SU(2)$,
we start from the parametrization
\bea
U &=& u_0 + i {\vec \s}\cdot {\vec u}\,,\nonumber\\
H &=& h_0 - i {\vec \s}\cdot {\vec h}\,,
\eea
where $u_\m$ are real and $h_\m$ complex. It is straightforward to compute
\bea
\Real\tr\{ UH\} &=& 2 (\Real h_0) u_0 + (\Real h_A)u_A  \nonumber\\
&=&  \Real\tr\{ U (\Real h_0) + i {\vec \s}\cdot (\Real {\vec h}) \}
\eea
We can then define an $SU(2)$ matrix 
$
V^\dagger = ((\Real h_0) + i {\vec \s}\cdot (\Real {\vec h}))/\rho
$
with
\be
\rho = \sqrt{(\Real h_0)^2+(\Real h_A)^2} \,,
\ee
in terms of which we can write
\be
u_2(H) = \int_{SU(2)} DU e^{\frac{1}{2}\Real\tr \{UH\}} =
         \int_{SU(2)} DU e^{\frac{\rho}{2} \tr \{U\}} \,.
\ee
The integral then is computed using the character expansion of the exponential
\be
e^{\frac{\rho}{2}\tr\{U\}} = \frac{2}{\rho}\sum_\n (2\n+1)I_{2\n+1}(\rho) \chi^{(\n)}(U)
\ee
and the result is
\be
u_2(H) = \frac{2}{\rho}I_1(\rho) \,.
\ee
In the mean-field we trade the 3 real degrees of freedom of $SU(2)$ (4 minus 1, from the determinant constraint)
for 4 independent mean-field degrees of freedom. The $SU(2)$ memory is encoded in the integral above with the
determinant constraint hidden in $\rho$.

\subsection{Boundary $U(1)$ links}

The $U(1)$ gauge group of the boundary links is 
embedded in $SU(2)$ as
\bea
U&=&e^{i \phi \s^3} = \cos(\phi) + i \sin(\phi)\,\sigma^3\,, \nonumber\\
H&=&h_0 - i h_3\,\sigma^3 \,.
\eea
The calculation proceeds like in the previous subsection
\be
u_1(H) =
\int_{U(1)\,\subset\,SU(2)} DU e^{\frac{1}{2}\Real\tr\{UH\}} = 
\frac{1}{2\pi}\int_0^{2\pi}d\phi e^{\rho \cos(\phi)} = I_0(\rho) \,,
\ee
where
\be
\rho = \sqrt{(\Real h_0)^2+(\Real h_3)^2} \,.
\ee

\section{The orbifold propagator \label{appb}}
 
The inverse mean-field propagator on the orbifold is
\be
{K} = -\left({K}^{(hh)}\right)^{-1}+({K}^{(vv)}+K^{({\rm gf})})\, .
\ee
As on the torus, $K^{(vh)}$ drops out from all expressions.
The propagator is written as
\be
K^{-1}=\Upsilon^{-1}K^{(hh)}, \hskip 1cm 
\Upsilon=-{\bf 1} + K^{(hh)}({K}^{(vv)}+K^{({\rm gf})})\label{Y}
\ee
so that it is the matrix $\Upsilon$ that is inverted when computing an observable.

We will Fourier transform along the four dimensions but not along the fifth dimension.
The propagator has then the components
\be
{K}^{-1} = {K}^{-1}(p',n'_5,M',\a' ;p'',n''_5,M'',\a'')
\ee
and in this Appendix we collect its pieces. 

\subsection{Fourier transformation to four-dimensional momentum space}

The Fourier transformation of a double derivative of an action $S$ 
\be
\frac{\d^2S}{\d v'\d v''} 
\ee
with respect to $v'\equiv v_{\a'}(n',M')$, $v''\equiv v_{\a''}(n'',M'')$ 
defines a kernel
\be
K(p',n_5',M',\a';p'',n_5'',M'',\a'') \,. \label{kernel}
\ee
The Fourier transformation is
\be
\frac{1}{{\cal N}^{(4)}} \sum_{\{n'_\m\},\{n''_\m\}}
{\rm e}^{ip'_\m n'_\m}\,{\rm e}^{-ip''_\m n''_\m}
{\rm e}^{i \frac{p'_{M'}}{2} (1-\d_{M'5})} \, 
{\rm e}^{-i \frac{p''_{M''}}{2} (1-\d_{M''5})} \, (-i)^{\d_{M'5}}\, (i)^{\d_{M''5}}
\frac{\d^2S}{\d v'\d v''} \,,
\ee
where ${\cal N}^{(4)}=TL^3$ is the four-dimensional volume.
The factors $(-i)^{\d_{M'5}}$ and $(i)^{\d_{M''5}}$ are introduced to make the
propagator real, a property that reflects CP invariance \cite{Narayanan:1995ex}.

The kernel \eq{kernel} is a matrix which is divided into sub-matrices in
coordinate indices $(n'_5,n''_5)$ for given $(p',M',\a';p'',M'',\a'')$.
These submatrices have dimension $(N_5+1)\times(N_5+1)$ if $M'=\m'$ and
$M''=\m''$; dimension $N_5\times(N_5+1)$ if $M'=5$ and
$M''=\m''$; dimension $N_5\times N_5$ if $M'=5$ and $M''=5$.

\subsection{$K^{(hh)}$ in four-dimensional momentum space}

$K^{(hh)}$ on the orbifold is defined as the second derivative 
\be
\frac{\partial^2}{\partial h_{\a'}(n',M')
\partial h_{\a''}(n'',M'')}
\ee
of the orbifold effective action. We introduce the notation
\bea
 a_2(\a,0)=a_2(\a,N_5)=\left\{\begin{array}{lcl} 
 0 & \mbox{if} & \a=1,2 \\
 a_2 & \mbox{if} & \a=3 \end{array}\right. 
\eea
and use the quantities
\bea
a_1 & = & \frac{1}{2} \left( \left.
\frac{u_1'(\rho)}{\rho}+\rho\left(\frac{u_1'(\rho)}{\rho}\right)^{\prime}
\right|_{\rho=\oh_0(0)=\oh_0(N_5)}\right) \,, \\
a_2 & = & \frac{1}{2} \left( \left. \frac{u_1'(\rho)}{\rho} \right|_{\rho=\oh_0(0)=\oh_0(N_5)}\right)
\,, \\
b_1(\cdot) & = & \left.
\frac{u_2'(\rho)}{\rho}+\rho\left(\frac{u_2'(\rho)}{\rho}\right)^{\prime}
\right|_{\rho=\oh_0(\cdot)} \,, \\
b_2(\cdot) & = & \left.
\frac{u_2'(\rho)}{\rho}\right|_{\rho=\oh_0(\cdot)} \,.
\eea
As mentioned, we perform a Fourier transformation along only the four-dimensional hyperplanes. 
The $\a'=\a''=0$ component is
\bea
{K}^{(hh), {\rm bulk}}_{00} &=& \dpp\delta_{M'M''}\d_{n_5'',n_5'} \Bigg\{
(1-\d_{M'5})\sum_{m_5=1}^{N_5-1}\d_{n_5',m_5} b_1(m_5)+\d_{M'5}\sum_{m_5=0}^{N_5-1}
\d_{n_5',m_5}b_1(m_5+1/2)\Bigg\}\nonumber\\ 
\label{Khh00bulk1}
\eea
and 
\bea
&&{K}^{(hh), {\rm bound.}}_{00} = \dpp\delta_{M'M''}\d_{n_5'',n_5'}
(1-\d_{M'5})\left(\d_{n_5',0}+\d_{n_5',N_5}\right)a_1 \label{Khh00bound}
\eea
The $\a'=\a''=A\ne 0$ components are
\bea
{K}^{(hh), {\rm bulk}}_{AA} &=& \dpp\delta_{M'M''}\d_{n_5'',n_5'} \Bigg\{
(1-\d_{M'5})\sum_{m_5=1}^{N_5-1}\d_{n_5',m_5}  b_2(m_5)+\d_{M'5}\sum_{m_5=0}^{N_5-1}
\d_{n_5',m_5} b_2(m_5+1/2)\Bigg\}\nonumber\\ 
\label{Khh00bulk2}
\eea 
and
\bea
{K}^{(hh), {\rm bound.}}_{AA} &=& \dpp\delta_{M'M''}\d_{n_5'',n_5'}
(1-\d_{M'5})\left (\d_{n_5',0} a_2(A,0) + \d_{n_5',N_5} a_2(A,N_5)\right)\label{KhhAAbound}
\eea

\subsection{$K^{(vv)}$ in four-dimensional momentum space}

\subsubsection{Gauge fixing}

We use the backward derivatives $f^A(n,M)=v^A(n,M)-v^A(n-\hat{M},M)$, $A=1,2,3$
and introduce the gauge fixing term
\bea
S_{\rm gf} &=&
\frac{1}{2\xi}\sum_{\{n_\m\}}\Biggl\{
\sum_{n_5=1}^{N_5-1}\sum_A\left[ \sum_{\m}f^A(n,\m) + \g f^A(n,5) \right]^2
\nonumber \\
&+& z_1\left[ 
\sum_{\m}f^3(\{n_\m\},n_5=0,\m) \right]^2
+ z_1\left[ 
\sum_{\m}f^3(\{n_\m\},n_5=N_5,\m) \right]^2
\Biggr\}\, . \label{Sgf}
\eea
We set the boundary weight to
\be
z_1 = \frac{1}{2}\,. \label{zs}
\ee
The Fourier transformation of ($v'\equiv
v_{A'}(n',M')$, $v''\equiv v_{A''}(n'',M'')$)
\be
\frac{\d^2S_{\rm gf}}{\d v'\d v''} 
\ee
we denote it by
\be
K^{({\rm gf})}(p',n_5',M',A';p'',n_5'',M'',A'')
\ee
and is divided into three contributions.

Contribution 1 is for $M'=\m'$, $M''=\m''$:
\bea
&&\frac{1}{\xi} \d^{(4)}_{p',p''}\hat{p'}_{\mu'}\hat{p'}_{\m''}\d_{n_5',n_5''}
\Biggl\{\d^{A',A''}\sum_A\d^{A',A}\sum_{n_5=1}^{N_5-1}\d_{n_5',n_5}
\nonumber\\
&+& z_1\d^{A',A''}\d^{A',3}\d_{n_5',0} + z_1\d^{A',A''} \d^{A',3}\d_{n_5',N_5} \Biggr\} \,,
\eea
where $\hat{p}_\m=2\sin(p_\m/2)$.

Contribution 2 is for $M'=5$, $M''=\m''$:
\bea
&&\frac{-\g}{\xi}\d^{(4)}_{p',p''}\hat{p'}_{\m''}
\d^{A',A''}\sum_A\d^{A',A}\sum_{n_5=1}^{N_5-1}
\left[\d_{n_5',n_5}\d_{n_5'',n_5'}-\d_{n_5',n_5-1}\d_{n_5'',n_5'+1}\right] \,.
\eea
The contribution for $M'=\m'$, $M''=5$ is obtained using the Hermiticity
property.

Contribution 3 is for $M'=5$, $M''=5$:
\bea
&&\frac{\g^2}{\xi}\d^{(4)}_{p',p''}
\d^{A',A''}\sum_A\d^{A',A}\sum_{n_5=1}^{N_5-1}\Box_{n_5',n_5''}(n_5) \,,
\eea
where
\be
\Box_{n_5',n_5''}(n_5) =
\d_{n_5',n_5''}(\d_{n_5',n_5}+\d_{n_5',n_5-1})
-\d_{n_5'',n_5'+1}\d_{n_5',n_5-1}-\d_{n_5'',n_5'-1}\d_{n_5',n_5} \,.
\ee
We have also implemented a background gauge fixing
\bea
f^A(n,\mu) & = & \ov_0(n_5)\left[v^A(n,\mu)-v^A(n-\hat{\mu},\mu)\right] \,,
\nonumber \\
f^A(n,5)   & = & \ov_0(n_5-\frac{1}{2})v^A(n,5)-
                 \ov_0(n_5+\frac{1}{2})v^A(n-\hat{5},5) \,,
\label{backgroundgf}
\eea
which is equally valid and vanishes when evaluated on the background $\ov_0$.
It does not change any of the results for physical observables.

\subsubsection{The Faddeev-Popov determinant}

Even though not directly relevant for the gauge propagator we can now carry out the 
Faddeev-Popov construction, necessary for the free energy.
The ghost action is
\be
S_{\rm FP} = 
\sum_{A',A''}\sum_{n',n''} {\overline c}^{A'}(n') 
{\cal M}^{({\rm orb})}_{A'n';A''n''} c^{A''}(n'')
\ee
where the sums run in the fundamental domain of the orbifold.
The ghost kernel
\be
{\cal M}^{({\rm orb})}_{A'n';A''n''} = \sum_M {\cal M}^{(M)}_{A'n';A''n''}
\ee
is obtained from the variation under infinitesimal gauge transformations
\be
\d f^{A'}(n,M) = \sum_{A'',n''} {\cal M}^{(M)}_{A'n;A''n''}
\omega^{A''}(n'')\, .
\ee
The changes on the orbifold with respect to the torus case in \cite{MFtorus}
is that on the boundaries the gauge transformation is $U(1)$ and acts only on 
$v^3$ at the boundaries at $n_5=0, N_5$. 

We define the matrices
\be
\d^{A',A''}\ov_0(n_5')
\left( 2\d_{n'',n'} - \d_{n'',n'+{\hat \n}} - \d_{n'',n'-{\hat \n}}\right)
\ee
with four-dimensional Fourier transformation
\be
{\cal M}^{(\n)}_{A'n'_5;A''n''_5} = \d^{A',A''}\ov_0(n_5')
\d^{(4)}_{p',p''}4\, \hat{p'}_\n^2\d_{n_5',n_5''}
\ee
and
\bea
{\cal M}^{(5)}_{A'n'_5;A''n''_5} &=& \d_{n_\n'',n_\n'} \Bigl[\d^{A'A''}\,
\left(\ov_0(n_5'+\frac{1}{2}) + \ov_0(n_5'- \frac{1}{2})\right)\,\d_{n_5'',n_5'} 
\nonumber \\
&-& \left[\d^{A',A''}+\d_{n_5',N_5-1}\d^{A',A''}(\d^{A',3}-1)\right]\,
\ov_0(n_5'+\frac{1}{2})\,\d_{n_5'',n_5'+{\hat 5}}
\nonumber \\
&-& \d^{A',A''}\,\left[1+\d_{n_5',1}(\d^{A',3}-1)\right]\,
\ov_0(n_5'-\frac{1}{2})\,\d_{n_5'',n_5'-{\hat 5}} \Bigr]
\eea
with the four-dimensional Fourier transformation changing
$ \d_{n_\n'',n_\n'} \to \d^{(4)}_{p',p''}$. The final expression for the Faddeev-Popov kernel is
\bea
{\cal M}^{({\rm orb})}_{A',n_5';A'',n_5''} &=& 
\sum_{l_5=1}^{N_5-1}
\d_{n_5',l_5}(1-\d_{l_5,0})(1-\d_{l_5,N_5})
\sum_\n \left[ {\cal M}^{(\n)}_{A',n'_5;A'',n''_5} + 
\g \, {\cal M}^{(5)}_{A',n'_5;A'',n''_5} \right]\nonumber\\
&+& z_1\sum_\n \left[ \d_{n_5',0} {\cal
    M}^{(\n)}_{3,n'_5;3,n''_5} + \d_{n_5',N_5}{\cal M}^{(\n)}_{3,n'_5;3,n''_5} 
\right]\,.
\eea
The relevant for the free energy Faddeev-Popov determinant is  
\be
D^{({\rm orb})}_{\rm FP} = {\rm det}\; {\cal M}^{({\rm orb})}.
\ee

\subsubsection{The double derivative of the plaquette action}

As for ${K}^{(hh)}$ due to the gauge invariance of the Wilson plaquette action 
the contributions of double derivatives with respect to links to the orbifold propagator
simplifies to
\bea
{K}^{(vv)}&=&\frac{\partial^2}{\partial v_{\a'}(n',M') \partial v_{\a''}(n'',M'')}
\Bigl[\sum_{n\in {\cal F}_0}\sum_{M} S_W[U(n,M] \Bigr]\nonumber\\
& +& \; {\rm boundary\; terms}
\eea
with ${\cal F}_0$ the fundamental domain of the orbifold with the boundary 
contributions separated out.

The gauge structure of the full kernel is
\bea
{K}^ {(vv)}_{\a'\a''}=\left(\begin{array}{cc}
(0,0) & 0\\
0 & (A',A'')\end{array}\right),
\eea
where we use the notation $(A',A'')$ for the blocks along the algebra indices.
Since
\bea
 \frac{\delta^2S_W}{\delta v'\,\delta v''} =
 \frac{\delta^2S_W}{\delta v''\,\delta v'} \,,
\eea
the kernel ${K}^ {(vv)}$ is Hermitian. Hence,
the off-diagonal elements are related by the Hermiticity relations
\bea
\left(K^{(vv)}_{\a'\a''}\right)_{M'M''} =
\left(\left(K^{(vv)}_{\a''\a'}\right)_{M''M'}\right)^\dagger
\eea
where $\left(K^{(vv)}_{\a'\a''}\right)_{M'M''}$ is a $(N_5+1)$-dimensional
matrix.

${K}^ {(vv)}$ depends on the following weights
\bea
v^{(\a',\a'')}_1(n_5) & = & \d_{\a',\a''}\, \ov_0(n_5+1/2) \ov_0(n_5+1) \,, \nonumber \\
v^{(\a',\a'')}_2(n_5) & = & \d_{\a',\a''}\, \ov_0(n_5)\ov_0(n_5+1/2) \,, \nonumber \\
u^{(\a',\a'')}_1(n_5) & = & v^{(\a',\a'')}_1(n_5) \,, \nonumber \\
u^{(\a',\a'')}_2(n_5) & = &  \d_{\a',\a''}\, \ov_0(n_5-1/2) \ov_0(n_5-1) \,. \label{weights1}
\eea
and
\bea
w^{(\a',\a'')}_1(n_5) & = &\d_{\a',\a''}\, (\ov_0(n_5))^2 \,, \nonumber \\
w^{(\a',\a'')}_2(n_5) & = & \d_{\a',\a''}\, \ov_0(n_5)\ov_0(n_5+1) \,, \nonumber \\
w^{(\a',\a'')}_3(n_5) & = & \d_{\a',\a''}\, (\ov_0(n_5+1/2))^2 \,, \nonumber \\
w^{(\a',\a'')}_4(n_5) & = & \d_{\a',\a''}\, (\ov_0(n_5-1/2))^2 \,.  \label{weights2}
\eea
The background is defined only for $\ov_0(n_5),\, n_5=0,1/2,1,3/2,\cdots , N_5$. 
Outside this range it vanishes by definition.
Correspondingly, the above definitions of the weights hold for every value of 
$n_5=0,\cdots,N_5$ in the range where the background is defined 
and for all values of $(\a',\a'')$ except from a few special cases which are
related to certain contributions from boundary or bulk/boundary plaquettes.
We specify these special cases below.

To begin, we set for $A'=A''=1$ and $A'=A''=2$ the weights that originate by
taking at least one derivative on one of the boundaries to zero:
\bea
w^{(1,1)}_1(0) &=& w^{(1,1)}_1(N_5) = 0 \,, \nonumber\\
w^{(1,1)}_4(N_5) &=& 0 \,, \nonumber\\
v^{(1,1)}_1(0) &=& 0\,, \nonumber\\
v^{(1,1)}_2(N_5-1) &=& 0\,, \nonumber\\
u^{(1,1)}_1(0) &=& u^{(1,1)}_2(N_5) = 0 \,, \nonumber\\
w^{(1,1)}_4(1) &=& w^{(1,1)}_3(N_5-1) = 0 \,, \nonumber\\
w^{(1,1)}_3(0)&=&0\,.
\eea
For the elements $A'=A''=3$ we have 
\bea
w_1^{(3,3)}(0) &=& 1/2\ov_0(0)^2 \, ,\nonumber\\
w^{(3,3)}_1(N_5) &=&  \frac{1}{2}\ov_0(N_5)^2 \,, \nonumber\\
w^{(3,3)}_4(N_5) &= & \ov_0(N_5-1/2)^2 \,, \nonumber\\
u^{(3,3)}_2(N_5) &= & \ov_0(N_5-1/2)\ov_0(N_5-1) \,, \nonumber\\
w^{(3,3)}_3(N_5-1) &= & \ov_0(N_5-1/2)^2 \, , \nonumber \\
v^{(3,3)}_2(N_5-1) &=& \ov_0(N_5-1)\ov_0(N_5-1/2) \,.
\eea
As mentioned, all other weights have their usual value, defined in \eq{weights1} and \eq{weights2}.
Weights that are not defined are assumed to be identically zero.
Finally, the anisotropy can be also absorbed in a redefinition of the
weights:
\bea
v^{(\a',\a'')}_{1,2} & \to & v_{1,2}\gamma \,, \nonumber\\
u^{(\a',\a'')}_{1,2} & \to & u_{1,2}\gamma \,,\nonumber\\
w^{(\a',\a'')}_1 & \to & w_1/\gamma \, , \nonumber\\
w^{(\a',\a'')}_{2,3,4} & \to & w_{2,3,4}\gamma 
\eea
for every $(\a',\a'')$.

We will now compute the various individual contributions to the full kernel. We define the symbols
\be
\epsilon(\a'; M',M'') = {\rm
  sgn}\left[\d_{M'M''}+(1-\d_{M'M''})(2\d_{\a',0}-1)\right] \label{epsi}
\ee
and 
\bea
&& 
{e}_{(p',n_5',M',\a' ; p'',n_5'',M'',\a'')} (N',N'';a,b,c,d) = \nonumber \\
&& \d^{(4)}_{p',p''} {\rm e}^{i \frac{p'_{M'}}{2} (1-\d_{M'5})} \, 
{\rm e}^{-i \frac{p''_{M''}}{2} (1-\d_{M''5})} \, (-i)^{\d_{M'5}}\, (i)^{\d_{M''5}}
\nonumber \\
&& \Bigl( a\, \d_{n_5'',n_5'} + 
\epsilon(\a'; M',M'')\, b\, \d_{n_5'',n_5'+\d_{N'5}}e^{-i p''_{N'}(1-\d_{N'5})} 
\nonumber\\
&& + \epsilon(\a'; M',M'')\, c\, \d_{n_5'',n_5'-\d_{N''5}}e^{i p'_{N''}(1-\d_{N''5})}
+ d\, \d_{n_5'',n_5'+\d_{N'5}-\d_{N''5}}
e^{-i p''_{N'}(1-\d_{N'5})} e^{i p'_{N''}(1-\d_{N''5})}\Bigr).\nonumber\\
\label{esymbol}
\eea
The double derivatives contribute then the components  
\bea
&& {K}^{(vv)}(x';x'') = -\b \cdot \Biggl[ \d_{M'M''}\d_{M'5}\sum_\m{e}_{(x';x'')} 
 \bigl(\m,\m; 0,w_2,w_2,0 \bigr)\nonumber\\
 &+&  \d_{M'M''}(1-\d_{M'5}) \Bigl(\sum_{\m\ne M'}{e}_{(x';x'')} 
 \bigl(\m,\m; 0,w_1,w_1,0 \bigr) + {e}_{(x';x'')} 
 \bigl(5,5; 0,w_3,w_4,0 \bigr)  \Bigr) \nonumber\\
 &+&  (1-\d_{M'M''}) (1-\d_{M'5})(1-\d_{M''5}){e}_{(x';x'')} 
 \bigl(M',M'';w_1,w_1,w_1,w_1 \bigr)\nonumber\\
 &+&  \d_{M'5}(1-\d_{M''5}){e}_{(x';x'')}
 \bigl(5,M'';v_1,v_2,v_1,v_2\bigr)\nonumber\\
 &+&  (1-\d_{M'5})\d_{M''5}{e}_{(x';x'')}
 \bigl(M',5;u_1,u_1,u_2,u_2 \bigr)\Biggr].
\eea
The argument of all weights in the above is $n_5'$ and their superscript
$(\a',\a'')$. Also we have defined the collective index 
$(x';x'')=(p',n_5',M',\a' ; p'',n_5'',M'',\a'')$.

To these components, the corresponding components of the
contributions from the gauge fixing term must be added, so that the full
contribution from the pure gauge part of the action to the propagator is
\be
K^{(vv)} + K^{({\rm gf})}\, .
\ee
We now have all the ingredients to compute $K^{-1}$ in \eq{Y}. The last issue
to be discussed before one does so is its eigenvalue structure.

A general property of $\Upsilon$ is that its eigenvalues are invariant under 
$p_\mu\longrightarrow  -p_\mu$
for any $\m=0,1,2,3$. This is useful when computing the free energy numerically.
Zero eigenvalues in $K^{(hh)}$, if any, clearly do not contribute anything to
the free energy or to any of the other observables.
The matrix $\Upsilon$ may have certain zero eigenvalues which on the other
hand must be taken care of otherwise it cannot be inverted. We separate these
zero modes in two  classes. 

One obtains spurious zero eigenvalues due to the Dirichlet boundary
conditions that force some of the fields to vanish on the boundaries.  For
example, for $\a'=\a''=1$ and $\a'=\a''=2$ the link variables $v_\a'(0,\m)$, $v_\a'(N_5,\m)$ all
vanish resulting into $2(d-1)$ ($d=5$) zero eigenvalues in the $11$ and $22$ components of
$\Upsilon$. These spurious eigenvalues occur for every value of
the four-dimensional momenta and appear just because bulk and boundary
components are packed together into the propagator. They are unphysical and
can be removed by hand.

The second class of zero eigenvalues of $\Upsilon$ contains those 
corresponding to vanishing four-dimensional momentum $p=0$.
On a finite lattice with fully periodic boundary
condition they appear in any gauge invariant formulation and persist even
after fixing completely the gauge, due to a left over global gauge invariance
surviving on a lattice \cite{DZ}. Beyond
the spurious eigenvalues mentioned above, the following properties hold for any $\alpha\neq0$ if
$p_\mu=0\;\forall\;\mu$:
without gauge fixing $\Upsilon$ has $N_5$ zero eigenvalues, corresponding to
local gauge transformations of the links along the extra dimension. After
gauge fixing through \eq{Sgf} (also in the variant of \eq{backgroundgf}), only
one zero mode survives. This is called ``toron''. If $p\neq0$ then there are
$N_5-1$ zero eigenvalue which are completely removed by the gauge fixing.
When present, such toron zero modes render
the corresponding physical observable, plagued by infinities, unusable. Since
however it is a finite volume effect, if (and only if) one can ensure the
existence of some regularization which leaves behind a finite result, they can
be dropped; their contribution to any physical observable is volume suppressed
and disappears when the infinite volume is taken \cite{MFtorus}. 

We finally report on the zero eigenvalues of the Faddeev-Popov determinant.
When the four dimensional momentum $p$ vanishes 
there are 2 zero eigenvalues for any $\alpha$. They correspond to lines of zeros 
in the Faddeev--Popov matrix ${\cal M}$. If $p$ is not zero there are two zero eigenvalues for $\alpha=1,2$ and none for $\alpha=3$.

\section{$Z$ and Higgs to leading order \label{s_appmass}}

We first define quantities that will be heavily used in the calculation of the observables. 
Let
\bea
\D_1^{(N_5)}(n_5) &=& \sum_{r=0}^{N_5-1}\frac{\d_{n_5, r}}{\ov_0(r+1/2)} =
(1-\d_{n_5,N_5})\frac{1}{\ov_0(n_5+\frac{1}{2})}
\eea
and
\bea
\Pi^{(1)}_{\langle 1,1\rangle}(\a,\b) &=& 
2\sum_{p_0'}\cos{p_0't} \sum_{n_5',n_5''} \D_1^{(N_5)}(n_5') 
{K}^{(-1)}(p_0',{\vec 0},n_5',5,\a;p_0',{\vec 0},n_5'',5,\b)
\D_1^{(N_5)}(n_5'')\nonumber\\ 
\Pi^{(2)}_{\langle 1,1\rangle}(\a,\b) &=& 
\sum_{p_0'}e^{ip_0't} \sum_{n_5',n_5''} \D_1^{(N_5)}(n_5') 
{K}^{(-1)}(p_0',{\vec p}',n_5',5,\a;p_0',{\vec p}',n_5'',5,\b)
\D_1^{(N_5)}(n_5'')\, .\nonumber\\ 
\label{Pisymbols}
\eea
We start from the $Z$ mass correlator. The computation is similar to
the one for the torus, so for more details see \cite{MFtorus}.
First we look at the single derivative of $P^{(0)}-P^{(0)\dagger}$ which
after the derivatives leaves behind the matrices
\be
\S^\a = 
\s^\a g {g}^\dagger + g {\overline \s}^\a {g}^\dagger - {\rm h.c.},
\ee
in the background. A simple calculation gives
\be
\S^0 = 0, \hskip .5cm \S^1 = 4i \s^1, \hskip .5cm \S^2 = 4i \s^2, \hskip .5cm \S^3 = 0 \, .
\ee
The group structure contributes terms of the form ${\rm tr}\{ {\overline \s}^3
\S^{\a_i\dagger} \S^{\a_j}\}$, 
to be contracted against the Euclidean structure, which we show below:
\bea
\frac{1}{L^6}\d_{{\vec p}',{\vec p}''}\d_{{\vec q}',{\vec q}''}\d_{{\vec p}',-{\vec q}''}
\eea
times
\bea
&& e^{i(p_l'-p_k')}\d^{\a_1A_1}\d^{\a_2A_2}\d^{\a_3A_3}\d^{\a_4A_4}K^{-1}(\a_1,\a_3)K^{-1}(\a_2,\a_4)\nonumber\\
&+& e^{i(p_l'+p_k')}\d^{\a_1A_2}\d^{\a_2A_1}\d^{\a_3A_3}\d^{\a_4A_4}K^{-1}(\a_1,\a_3)K^{-1}(\a_2,\a_4)\nonumber\\
&+& e^{-i(p_l'+p_k')}\d^{\a_1A_1}\d^{\a_2A_2}\d^{\a_4A_3}\d^{\a_3A_4}K^{-1}(\a_1,\a_3)K^{-1}(\a_2,\a_4)\nonumber\\
&+& e^{-i(p_l'-p_k')}\d^{\a_1A_2}\d^{\a_2A_1}\d^{\a_3A_4}\d^{\a_4A_3}K^{-1}(\a_1,\a_3)K^{-1}(\a_2,\a_4)\nonumber\\
&+& e^{i(p_l'-p_k')}\d^{\a_1A_1}\d^{\a_2A_2}\d^{\a_3A_4}\d^{\a_4A_3}K^{-1}(\a_1,\a_4)K^{-1}(\a_2,\a_3)\nonumber\\
&+& e^{i(p_l'+p_k')}\d^{\a_1A_2}\d^{\a_2A_1}\d^{\a_3A_4}\d^{\a_4A_3}K^{-1}(\a_1,\a_4)K^{-1}(\a_2,\a_3)\nonumber\\
&+& e^{-i(p_l'+p_k')}\d^{\a_1A_1}\d^{\a_2A_2}\d^{\a_3A_3}\d^{\a_4A_4}K^{-1}(\a_1,\a_4)K^{-1}(\a_2,\a_3)\nonumber\\
&+& e^{-i(p_l'-p_k')}\d^{\a_1A_2}\d^{\a_2A_1}\d^{\a_3A_3}\d^{\a_4A_4}K^{-1}(\a_1,\a_4)K^{-1}(\a_2,\a_3)\, .
\eea
Putting everything together, the observable can be first written as
\bea
&& C_{Z}^{(2)}(t)=\frac{c^{(2)}}{4!}(P_0^{(0)})^4 (v_0(0))^4\frac{1}{{({\cal N}^{(4)}})^2}\sum_{A_1,\cdots, A_4} 
{\rm tr}\{ {\overline \s}^3 \S^{A_1\dagger} \S^{A_2}  \}
{\rm tr}\{ {\overline \s}^3 \S^{A_3\dagger} \S^{A_4}  \}\cdot 2  \cdot 
\sum_{{\vec p}'}\sum_{k,l}\Biggl\{ \nonumber\\
&& e^{i(p_l'-p_k')} 
\Pi^{(2)}_{\langle 1,1\rangle}(A_1,A_3)\Pi^{(2)}_{\langle 1,1\rangle}(A_2,A_4)
+ e^{-i(p_l'-p_k')} 
\Pi^{(2)}_{\langle 1,1\rangle}(A_2,A_4)\Pi^{(2)}_{\langle 1,1\rangle}(A_1,A_3)\nonumber\\
&+& e^{i(p_l'+p_k')} 
\Pi^{(2)}_{\langle 1,1\rangle}(A_2,A_3)\Pi^{(2)}_{\langle 1,1\rangle}(A_1,A_4)
+ e^{-i(p_l'+p_k')} 
\Pi^{(2)}_{\langle 1,1\rangle}(A_1,A_4)\Pi^{(2)}_{\langle 1,1\rangle}(A_2,A_3)\Biggr\}
\eea
and finally be simplified to
\be
C_{Z}^{(2)}(t)=\frac{c^{(2)}\cdot 4\cdot 4096}{24{({\cal N}^{(4)}})^2}(P_0^{(0)})^4 (v_0(0))^4
\sum_{\vec p'}\sum_k\sin^2{p_k'}\cdot
\Pi^{(2)}_{\langle 1,1\rangle}(1,1) \Pi^{(2)}_{\langle 1,1\rangle}(2,2). \label{CZapp}
\ee
We denote by $P_0^{(0)}$ the Polyakov loop evaluated on the background.
In the above we have used that the elements of the propagator which are
diagonal in the $M$ index are invariant under ${\vec p}'\to
-{\vec p}'$. The coefficient $c^{(2)}$ is the symmetry factor of the double
exchange diagram and it is
\be
c^{(2)} = \left(\begin{array}{c} 4\\ 2 \end{array}\right).
\ee
We note that the toron $(p_\mu=0\;\forall\mu)$ does not contribute to \eq{CZapp}.

Next we turn to the Higgs mass. The relevant trace is now
\be
\langle 1 \rangle_\a = {\rm tr}\{ \s^\a g g^\dagger +
g{\overline \s}^{\a}g^\dagger \}.
\ee
Apart from the traces originating from the above structure,
the computation is again analogous to the one for the torus geometry (see \cite{MFtorus}) so we give directly the result:
\be
C_H^{(1)}(t) = \frac{1}{{\cal N}^{(4)}}(P_0^{(0)})^2 \frac{1}{2!}c^{(1)}
\sum_\alpha \langle 1 \rangle_\a^2 \Pi^{(1)}_{\langle 1,1\rangle}(\a,\a),
\ee
summation over repeated gauge indices implied. The symmetry factor is
\be
c^{(1)} = 1\, .
\ee
Performing the traces one obtains
\be
C_H^{(1)}(t) = \frac{1}{{\cal N}^{(4)}}(P_0^{(0)})^2 \frac{16}{2!} \Pi^{(1)}_{\langle 1,1\rangle}(0,0).
\ee

\end{appendix}



\begin{thebibliography}{99}

\bibitem{MFtorus}
N. Irges and F. Knechtli,\\
\emph{Mean-field gauge interactions in five dimensions I. The Torus.}\\
Nucl.\ Phys.\  {\bf B822} (2009) 1. arXiv:0905.2757
[hep-lat]. Erratum-ibid.{\bf B840} (2010) 438. 

\bibitem{DZ}
J.~M.~Drouffe and J.~B.~Zuber,\\
\emph{Strong Coupling And Mean Field Methods In Lattice Gauge Theories},\\
Phys.\ Rept.\  {\bf 102} (1983) 1.

\bibitem{NFconfinement}
N. Irges and F. Knechtli,\\
\emph{A new model for confinement},\\
arXiv:0910.5427 [hep-lat].

\bibitem{HMR}
A. Hebecker and J. March-Russell,\\
\emph{ The structure of GUT breaking by orbifolding},\\
Nucl.\ Phys.\  {\bf B625} (2002) 128. hep-ph/0107039.

\bibitem{Hosotani}
N. S. Manton,\\
\emph{A New Six-Dimensional Approach to the Weinberg-Salam Model},\\
Nucl.\ Phys.\  {\bf B158} (1979) 141.\\
Y. Hosotani,\\
\emph{Dynamical Gauge Symmetry Breaking as the Casimir Effect},\\
Phys.\ Lett.\  {\bf B129} (1983) 193.

\bibitem{5Dan}
Y. K. Fu and H. B. Nielsen,\\
\emph{A Layer phase in a nonisotropic $U(1)$ lattice 
gauge theory: dimensional reduction, a new way},\\
Nucl.\ Phys.\  {\bf B236} (1984) 167.\\
D. Berman and E. Rabinovici,\\
\emph{Layer phases in anisotropic lattice gauge theories},\\
Phys.\ Lett.\  {\bf B66} (1985) 292.

\bibitem{5DFP}
H. Gies,\\
\emph{Renormalizability of gauge theories in extra dimensions},\\
Phys.\ Rev.\ {\bf D68} (2003) 085015. hep-th/0305208.\\
T.R. Morris,\\
\emph{Renormalizable extra-dimensional models},\\
JHEP {\bf 0501} (2005) 002. hep-ph/0410142.

\bibitem{LPMC}
A. Huselbos, C. P. Korthals-Altes and S. Nicolis,\\
\emph{Gauge theories with a layered phase},\\
Nucl.\ Phys.\  {\bf B450} (1995) 437. hep-th/9406003. \\
P. Dimopoulos, K. Farakos, A. Kehagias and G. Koutsoumbas,\\
\emph{Lattice evidence for gauge field localization on a brane},\\
Nucl.\ Phys.\  {\bf B617} (2001) 237. hep-th/0007079.\\
P. Dimopoulos, K. Farakos and G. Koutsoumbas,\\
\emph{The phase diagram for the anisotropic SU(2) adjoint Higgs
model in  5D: Lattice evidence for layered structure},\\
Phys.\ Rev.\ {\bf D65} (2002) 074505. hep-lat/0111047. \\
K. Farakos, P. de Forcrand, C. P. Korthals-Altes, M. Laine and
M. Vettorazzo,\\
\emph{Finite temperature Z(N) phase transition with Kaluza-Klein gauge
  fields},\\
Nucl.\ Phys.\  {\bf B655} (2003) 170. hep-ph/0207343.\\
P. Dimopoulos, K. Farakos and S. Vrentzos,\\
\emph{The 4-D layer phase as a gauge field localization: 
Extensive study of  the 5-D 
anisotropic U(1) gauge model on the lattice},\\
Phys.\ Rev.\ {\bf D74} (2006) 094506. hep-lat/0607033. \\
K. Farakos and S. Vrentzos,\\
\emph{Establishment of the Coulomb law in the layer phase of a
pure U(1) lattice gauge theory},\\
Phys.\ Rev.\ {\bf D77} (2008) 094511. arXiv:0801.3722 [hep-lat]. 

\bibitem{1PTMC}
M.~Creutz,\\
\emph{Confinement and the criticality of space-time},\\
Phys.\ Rev.\  Lett.\ {\bf 43} (1979) 553.

\bibitem{2PTMC}
S. Ejiri, J. Kubo and M. Murata,\\
\emph{A study on the nonperturbative existence of 
Yang-Mills theories with large extra dimensions},\\ 
Phys.\ Rev.\  {\bf D62} (2000) 105025. hep-ph/0006217.\\
P. de Forcrand, A. Kurkela and M. Panero,\\
\emph{The phase diagram of Yang-Mills theory with a compact extra dimension},\\
JHEP 1006 (2010) 050.\\
K. Farakos and S. Vrentzos,\\
\emph{Exploration of the phase diagram of 5d anisotropic SU(2) gauge theory},\\
Nucl.\ Phys.\ B {\bf 862} (2012) 633. arXiv:1007.4442 [hep-lat].\\
L. Del Debbio, A. Hart and E. Rinaldi,\\
\emph{Light scalars in strongly-coupled extra-dimensional theories},\\
arXiv:1203.2116 [hep-lat].

\bibitem{FrancAM}
F. Knechtli, M. Luz and A. Rago,\\
\emph{On the phase structure of five-dimensional SU(2) gauge theories with anisotropic couplings},\\
Nucl.\ Phys.\  {\bf B856} (2012) 283. arXiv:1110.4210 [hep-lat].

\bibitem{NFMC}
N. Irges and F. Knechtli,\\
\emph{Non-perturbative mass spectrum of an extra-dimensional orbifold},\\
hep-lat/0604006.\\
N. Irges and F. Knechtli,\\
\emph{Lattice gauge theory approach to spontaneous symmetry
breaking from an extra dimension},\\
Nucl.\ Phys.\  {\bf B775} (2007) 283. hep-lat/0609045.

\bibitem{MM}
K. Ishiyama, M. Murata, H. So and K. Takenaga,\\
\emph{Symmetry and $\mathbb{Z}_2$ orbifolding approach in five-dimensional lattice gauge theory},\\
Prog. Theor. Phys. {\bf 123} (2010) 257.

\bibitem{ABQ}
I. Antoniadis, K. Benakli and M. Quiros,\\
\emph{Finite Higgs mass without supersymmetry},\\
New\ J.\ Phys.\  {\bf 3} (2001) 20. hep-th/0108005.

\bibitem{OrbPert}
G. von Gersdorff, N. Irges and M. Quiros,\\
\emph{Bulk and brane radiative effects in gauge theories on orbifolds},\\
Nucl.\ Phys.\  {\bf B635} (2002) 127. hep-th/0204223.\\
H-C. Cheng, K. Matchev and M. Schmaltz,\\
\emph{Radiative corrections to Kaluza-Klein masses},\\
Phys.\ Rev.\  {\bf D66} (2002) 036005. hep-ph/0204342.

\bibitem{NFM}
N. Irges, F. Knechtli and M. Luz,\\
\emph{The Higgs mechanism as a cut-off effect},\\
JHEP {\bf 08} (2007) 028. arXiv:0706.3806 [hep-ph].\\
F.~Knechtli, N.~Irges and M.~Luz,\\
\emph{New Higgs mechanism from the lattice},\\
J.\ Phys.\ Conf.\ Ser.\  {\bf 110} (2008) 102006.
[arXiv:0711.2931 [hep-ph]].

\bibitem{Symanzik}
K. Symanzik,\\
\emph{Some topics in quantum field theory},\\
Math.\ Prob.\ Theor.\ Phys.\ {\bf 153} (1982) 47.

\bibitem{Kubo:2001zc}
M. Kubo, C. S. Lim and H. Yamashita,\\
\emph{The Hosotani mechanism in bulk gauge theories with an
orbifold extra space $S^1/\mathbb{Z}_2$},\\
Mod.\ Phys.\  Lett.\ {\bf A17} (2002) 2249. hep-ph/0111327.

\bibitem{SSS}
C. Scrucca, M. Serone and L. Sivestrini,\\
\emph{Electroweak symmetry breaking and fermion masses from
extra dimensions},\\
Nucl.\ Phys.\  {\bf B669} (2003) 128. hep-ph/0304220.

\bibitem{Ruhl:1982er}
W.~R\"uhl,\\
\emph{The Mean Field Perturbation Theory Of Lattice Gauge Models With Covarian
t Gauge Fixing},\\
Z.\ Phys.\ {\bf C18} (1983) 207.

\bibitem{NForbdef}
N. Irges and F. Knechtli,\\
\emph{Non-perturbative definition of five-dimensional gauge
theories on the $\mathbb{R}^4\times S^1/\mathbb{Z}_2$ orbifold},\\
Nucl.\ Phys.\  {\bf B719} (2005) 121. hep-lat/0411018.

\bibitem{Knechtli:2005dw}
F. Knechtli, B. Bunk and N. Irges,\\
\emph{Gauge theories on a five-dimensional orbifold},\\
PoS\  {\bf LAT2005} (2006) 280.

\bibitem{Narayanan:1995ex}
  R.~Narayanan and U.~Wolff,\\
\emph{Two loop computation of a running coupling in lattice Yang-Mills theory},\\
Nucl.\ Phys.\  B {\bf 444} (1995) 425. hep-lat/9502021.

\end{thebibliography}
\end{document}